\documentclass[journal]{vgtc}                
\ifpdf
  \pdfoutput=1\relax                   
  \usepackage{graphicx}                
  \DeclareGraphicsExtensions{.pdf,.png,.jpg,.jpeg} 
\else
  \usepackage{graphicx}                
  \DeclareGraphicsExtensions{.eps}     
\fi%

\graphicspath{{figures/}{pictures/}{images/}{./}} 
\newcommand\numberthis{\addtocounter{equation}{1}\tag{\theequation}}
\usepackage{microtype}                 
\PassOptionsToPackage{warn}{textcomp}  
\usepackage{textcomp}                  
\usepackage{mathptmx}                  
\usepackage{times}                     
\usepackage{cite}                      
\usepackage{tabu}                      
\usepackage{booktabs}                  

\usepackage{color}
\usepackage{graphics,url,subfigure}
\usepackage{multirow}
\usepackage{float}
\usepackage{amsfonts,amssymb,euscript,mathrsfs,bbm,theorem}
\usepackage{amsmath,scalefnt}
\usepackage{algorithm}
\usepackage{algorithmic}
\usepackage{verbatim}
\usepackage{graphicx,dblfloatfix}
\usepackage[labelsep=period]{caption}
\usepackage{theorem}
\DeclareGraphicsExtensions{.ps}

\definecolor{mycolor}{rgb}{0.00, 0.0, 0.0}
\definecolor{mycolor2}{rgb}{0, 0.5, 0}
\definecolor{mycolor3}{rgb}{0, 0, 0.5}
\definecolor{mycolor4}{rgb}{1, 0, 0}

\newcommand{\ellie}[1]{{\color{mycolor}{#1}}}
\newcommand{\ta}[1]{{\color{mycolor}{#1}}}
\newcommand{\tma}[1]{{\color{mycolor}{#1}}}
\newcommand{\delete}[1]{\iffalse{#1}\fi}

\newcommand{\etal}{\textit{et\,al.}}

\newcommand{\vx}{\vect{x}}

\newcommand{\vv}{\vect{v}}

\newcommand{\vect}[1]{\boldsymbol{\mathbf #1}}

\newcommand{\norm}[1]{\lVert#1\rVert}

\newcommand{\transp}[1]{{#1}^{\mathrm T}}

\newcommand{\reel}{\mathbb{R}}
\newcommand{\rthree}{\reel^3}

\newcommand{\SC}[1]{Sec.~\ref{#1}}

\newcommand{\EQ}[1]{(\ref{#1})}

\newcommand{\FG}[1]{Fig.~\ref{#1}}

\newcommand{\bxi}{{\pmb \xi}}

\DeclareMathOperator{\grad}{\nabla}

\newcommand{\shah}{{\textstyle \amalg{\kern-4.pt\amalg}}}

\usepackage{tabularx}

\onlineid{1130}

\vgtccategory{Research}
\vgtcpapertype{algorithm/technique}

\title{Direct Volume Rendering with Nonparametric Models\\ of Uncertainty}

\author{Tushar Athawale, Bo Ma, Elham Sakhaee, Chris R. Johnson,~\textit{Fellow,~IEEE,}\\ and Alireza Entezari,~\textit{Senior Member,~IEEE}}
\authorfooter{
\item
 T. Athawale and C. R. Johnson are with Scientific Computing \& Imaging (SCI) Institute, University of Utah. E-mail:  \{tushar.athawale, crj\}@sci.utah.edu,
\item
B. Ma, E. Sakhaee, and A. Entezari are with the Department of CISE at the University of Florida, Gainesville, FL, 32611. E-mail: \{bbo, esakhaee, and entezari\}@cise.ufl.edu
}

\shortauthortitle{Biv \MakeLowercase{\textit{et al.}}: Global Illumination for Fun and Profit}

\abstract{We present a nonparametric statistical framework for the quantification, analysis, and propagation of data uncertainty in direct volume rendering (DVR). The state-of-the-art statistical DVR framework allows for preserving the transfer function (TF) of the ground truth function when visualizing uncertain data; however, the existing framework is restricted to parametric models of uncertainty. In this paper, we address the limitations of the existing DVR framework by extending the DVR framework for nonparametric distributions. We exploit the \emph{quantile interpolation} technique to derive probability distributions representing uncertainty in viewing-ray sample intensities in closed form, which allows for accurate and efficient computation. \tma{We evaluate our proposed nonparametric statistical models through qualitative and quantitative comparisons with the mean-field and parametric statistical models, such as uniform and Gaussian, as well as Gaussian mixtures.} In addition, we present an extension of the state-of-the-art rendering parametric framework to 2D TFs for improved DVR classifications. We show the applicability of our uncertainty quantification framework to ensemble, downsampled, and bivariate versions of scalar field datasets. } 


\CCScatlist{ 
 \CCScat{K.6.1}{Management of Computing and Information Systems}%
{Project and People Management}{Life Cycle};
 \CCScat{K.7.m}{The Computing Profession}{Miscellaneous}{Ethics}
}

\teaser{
  \centering
  \includegraphics[width=\linewidth]{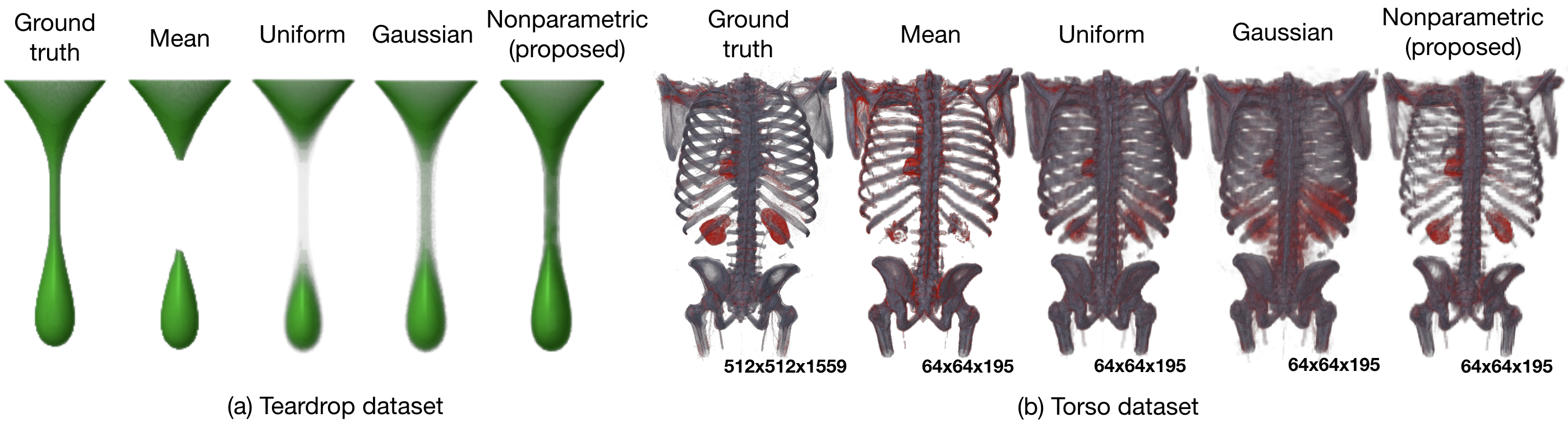}
 \caption{Nonparametric models of uncertainty improve the quality of reconstruction and classification within an uncertainty-aware direct volume rendering framework. (a) Improvements in topology of an isosurface in the teardrop dataset ($64 \times 64 \times 64$) with uncertainty due to sampling and quantization. (b) Improvements in classification (i.e., bones in gray and kidneys in red) of the torso dataset with uncertainty due to downsampling.}
  \label{fig:teaser}
}
 
\vgtcinsertpkg

\begin{document}


\firstsection{Introduction}

\maketitle

As visualization techniques continue to facilitate the exploration of scientific simulations and biomedical datasets, analysis of data uncertainties, inherent in all forms of acquisition, modeling, and representation, has emerged as an important research area. Uncertainties present in data, such as those intrinsic to acquisition or modeling (e.g., sampling, quantization), as well as those introduced within data processing (e.g., filtering/downsampling), adversely impact the reliability of visualizations. To facilitate reliable visualization in the presence of uncertainty, several studies have advocated redesigning visualization algorithms to treat data as probability distributions to account for various types of uncertainty~\cite{wong2012top,uncertainty}. Quantifying the impact of uncertainty in the computational process and its propagation throughout the visualization pipeline poses several mathematical as well as algorithmic challenges. 

Visualization of uncertain data is an active field of research, including several advances in innovative ways for the visual depiction of uncertainty~\cite{brodlie2012review}. In contrast, analysis and propagation of uncertainty in the various stages of the rendering pipeline and quantifying their impact on transforming the uncertainty remain challenging tasks. Recent studies~\cite{fout2012FVR, kronander2010estimation, Correa2009framework} have considered sources of uncertainty within the visual analytics process  and analyzed the contribution of each stage to the uncertainty associated with the volume data. 


In our work, we study the propagation of data uncertainty through the stages of the \ta{direct volume rendering} (DVR) pipeline. DVR is a fundamental visualization technique for gaining insights into volumetric datasets. A \ta{transfer function} (TF) plays a central role in DVR, as it translates scalar or multifield data to optical properties, such as color and opacity. \ta{The visual mappings produced by TFs help users understand interesting features or patterns in the dataset. Such a process of feature identification through TF space exploration is referred to as \emph{classification}~\cite{Hadwiger06}.} 


\ta{The classification task in} DVR can be challenging when the data have uncertainty. The DVR of uncertain data by reusing the TF design for the original function can lead to poor classification results~\cite{younesy2006improving}.  A simple workaround would be to generate a new TF when data have uncertainty; but the design of TFs is known to be a time-consuming and laborious task, especially in the case of multidimensional TFs. Thus, developing new techniques that seek to improve the quality of visualizations of uncertain data while reusing the TF design for the original function is more desirable. An expressive rendering of uncertain data must preserve all visible features in the rendering of the ground truth data and also indicate the uncertainties engendered by the various stages of the rendering pipeline. 

A recently developed statistical framework by Sakhaee \etal~\cite{Entezari2016Statistical} introduced a novel approach for DVR that addressed the issue of preserving TF designs of the original function for visualizations of uncertain data. In their approach, data uncertainty is integrated against 1D TF in the reconstruction stage of the DVR pipeline. Their framework opens up new directions for the exploration of uncertain data, because it allows for uncertainty propagation and aggregation within the reconstruction and the traditional classification stages. In the framework proposed by Sakhaee \etal~\cite{Entezari2016Statistical}, the data input into the DVR process are considered as a field of random variables described by parametric probability density functions (\ta{PDFs}). Liu \etal~\cite{TA:2012:LLBP} proposed a DVR framework for visualization of uncertain 3D data when {\ta{PDFs}} are modeled using Gaussian mixture models (GMM). Their framework used an expensive Monte Carlo \ta{(MC)} sampling approach for uncertainty estimation of interpolated samples of a DVR raycaster.

Inspired by contributions on DVR for parametric-~\cite{Entezari2016Statistical} and GMM-~\cite{TA:2012:LLBP} based uncertainty, we propose a closed-form DVR framework for nonparametric density models. Recently, noise modeling using nonparametric distributions has been advocated over parametric distributions for taking into account the skewness or multimodality of distributions, and hence improving the precision of uncertainty visualizations~\cite{pothkow2013nonparametric, AE15}. Although the extension to nonparametric distributions for DVR has been discussed in previous work~\cite{Entezari2016Statistical}, such an extension is challenging, especially from the computational cost point of view, and no recipe for implementation or empirical results has been provided. 

In our work, we present an efficient quantile interpolation technique for DVR of uncertain data, where uncertainty is characterized using nonparametric distributions. Read~\cite{TA:1999:Read} first introduced the $1$D quantile interpolation technique for the interpolation of histograms. Hollister and Pang~\cite{TA:2015:HP} leveraged the quantile interpolation technique for bilinear interpolation of nonparametric distributions characterizing uncertain vector fields. We present quantile interpolation for trilinear interpolation of nonparametric distributions, and we successfully integrate interpolated distributions with a DVR framework for the visualization of uncertain data.



Although the-state-of-the-art spline-based technique~\cite{Entezari2016Statistical} explores $1$D classification of uncertain scalar fields for DVR, the classification of uncertain scalar volumes with multidimensional transfer functions (TFs) and the visualization of multifield data remain challenging tasks. Specifically, the intensity-gradient magnitude ($2$D) TF has proved valuable due to its effectiveness in isolating complex boundaries with overlapping materials~\cite{kniss2002multidimensional}. Unlike the previous work that dealt with uncertainties in the data and gradient field for data separately, we leverage the simultaneous estimation of uncertainty in both the scalar and gradient fields. Specifically, we apply a spline-based statistical framework to intensity-gradient magnitude 2D TFs and study its ramifications in visualizing bivariate datasets.



\subsection{Contributions} 
We generalize the recently developed spline-based statistical framework~\cite{Entezari2016Statistical} to nonparametric statistical models and 2D TFs for visualization of uncertain data. Specifically, we propose the following methods for the visualization of uncertain data:

\begin{itemize}
\item Given an uncertain scalar field, represented as a field of \ta{PDFs}, we analytically derive the interpolation of the  \ta{PDFs} of the intensities for any arbitrary sample point along the viewing rays for DVR. Each grid point is modeled using a nonparametric \ta{PDF} in contrast to a parametric one. A previous study~\cite{Entezari2016Statistical} considered nonparametric models only as a possible venue for investigation within
the framework for uncertainty visualization, but did not explore
nonparametric models or their potential advantages. In our work, we fill this gap by proposing the use of the quantile interpolation technique for nonparametric statistics. Specifically, we present an analytic formulation of the quantile interpolation technique for trilinear interpolation of nonparametric \ta{PDFs}. Our closed-form formulation permits efficient integration of nonparametric statistics with a DVR framework. \ta{We demonstrate the effectiveness of our proposed nonparametric models through qualitative and quantitative comparisons with mean-field and parametric models.} 

\item \ta{The quantile interpolation technique presents an example of order statistics, where quantiles are ordered using a cumulative density function for a random variable. We, thus, take advantage of order statistics to investigate uncertainty in ensemble data by devising a tool called the quartile view.}


\item Similar to intensities, we analytically derive a formulation for the interpolation of \ta{PDFs} of the gradient magnitudes for samples along the viewing rays. The \ta{PDFs} of intensities and gradient magnitudes are then integrated against 2D TF (gradient magnitude vs. intensity). Improved classification of uncertain scalar fields using this approach signifies the importance of the simultaneous handling of uncertainty in data and its gradient field. 

\item We demonstrate an application of our proposed DVR framework for the visualization of ensemble and downsampled data. We also present an application of the reconstruction of \ta{PDFs} for DVR of bivariate data.
 
\end{itemize}


The paper is organized as follows: In \SC{sc:LitReview}, we review the prior work on uncertainty visualization and multidimensional TFs. 
In \SC{sec:stateOftheArt}, we briefly revisit the state-of-the-art theory~\cite{Entezari2016Statistical} on the interpolation of uncertain scalar fields and linear interpolation of histograms in 1D using the quantile interpolation technique~\cite{TA:1999:Read}. We then present an extension of the quantile interpolation technique for trilinear interpolation of nonparametric PDFs and its integration into a DVR framework in \SC{sec:3dQuantileInterpolation} and \SC{sec:dvrQuantile}, respectively. \ta{In \SC{sec:quartileView}, we describe our quartile view technique.} In \SC{sec:gradientInterp} and \SC{sec:ProbTF}, we describe a spline-based model for the interpolation of uncertain gradient fields and propose integration of interpolated intensity and gradient magnitude distributions against 2D TFs for visualizations. In \SC{sec:exp}, we demonstrate experimental results for the visualization of uncertain data using reconstructed uncertain scalar fields and uncertain gradient fields. Finally, we conclude our work and discuss possible future work in \SC{concl}.

\section{Related Work}\label{sc:LitReview}
\subsection{Uncertainty Visualization} 
Uncertainty visualization has been recognized as one of the top challenges in the visualization community due 
to its significance in decision-making~\cite{uncertainty, top-scivis, zuk2006theoretical, wong2012top}. Specifically, uncertainty visualization has been to shown to be important in avoiding misleading interpretations regarding the underlying data. Whereas classical visualization approaches consider uncertainty associated with the volume data~\cite{pang1997approaches,potter2012quantification}, 
recent works account for aggregated uncertainty due to rendering algorithms~\cite{fout2012FVR,Correa2009framework, Entezari2016Statistical}. Brodlie \etal~\cite{brodlie2012review} discuss the impact of propagating uncertainty in the data to uncertainty in the final image. They define the propagation problem as determining the \ta{PDF} of the output entities from the \ta{PDF} of the input entities, or discuss that often a \ta{MC sampling method} is required to obtain the \ta{PDF} of the output. In this paper, we derive \ta{PDFs} analytically. Correa \etal~\cite{Correa2009framework} describe uncertainty propagation and aggregation 
for data transformations, such as regression, principal component analysis, and k-means clustering. 

Statistical uncertainty analysis of topological features of data, such as level sets and critical points, has drawn increasing attention in the study of data uncertainty. Contour~\cite{SCI:Whi2013a}, curve~\cite{SCI:Mir2014a}, and surface~\cite{SCI:Gen2014a} boxplots extend the concept of functional-depth ranking~\cite{pintado2009fd} for deriving quantiles that represent the spatial variability of ensembles of isocontours, arbitrary curves, and 2D images, respectively. The level-crossing probability method of P\"othkow and Hege~\cite{pothkow2011positional, pothkow2013nonparametric} and uncertainty-aware marching cubes algorithm proposed by Athawale \etal~\cite{Athawale:2013:UQL, AE15, Athawale:2018:PAD} demonstrated the benefits of nonparametric statistical noise modeling over parametric modeling for deriving positional uncertainty in level sets. Hixels~\cite{Thompson:2013:LDAV} summarized information of a brick of volume as a histogram for visualizing fuzzy isosurfaces. Suter \etal~\cite{suter2014visual} exploited shape similarities based on Hausdorff distance for extracting isosurfaces from 3D scalar fields. G{\"u}nther \etal~\cite{TA:2014:DSJ} and Favelier \etal~\cite{TA:2018:PA} devised statistical approaches for characterizing spatial variations in critical points of uncertain data. Otto~\etal~performed statistical gradient field analysis for visualizing topological variations of uncertain 2D~\cite{otto2010uncertain} and 3D~\cite{otto2011uncertain} vector fields. Recently, He \etal~\cite{TA:2018:HGS} devised a nonparametric method, known as surface density estimation (SDE), for analyzing spatial inconsistency in level sets. 



In the context of DVR, a considerable body of literature has analyzed uncertainty propagation in the DVR pipeline and the impact of errors on final visualizations. Fout and Ma~\cite{fout2012FVR} discussed the contribution of each stage of DVR (quantization, reconstruction/filtering, classification, shading, and integration) to the uncertainty associated with volume data. Kniss \etal~\cite{kniss:2005:SQVV} presented rendering based on probabilistic classification that allows the user to interactively explore the uncertainty and the information computed during fuzzy segmentation. Pfaffelmoser \etal~\cite{TA:2011:PRW} assumed Gaussian-distributed data uncertainty for visualizing geometric uncertainty in isosurfaces extracted using DVR. Etian \etal~\cite{etiene2014verifying, SCI:Eti2015b} proposed a novel strategy for verifying the correctness of DVR implementations by analyzing the correlation between discretization errors caused by sampling along viewing rays and the rendering quality. Kronander \etal~\cite{kronander2010estimation} evaluated the effects of the propagation of numerical errors, caused by finite precision of data representation and processing, on the volume rendered images. Djurcilov \etal~\cite{Djurcilov02} employed features such as speckle, texture, or noise to represent uncertainty in the volume rendering process. In medical volume rendering, probabilistic animation has been used to visualize uncertainty~\cite{lundstrom2007uncertainty}. 

The focus of our work is on advocating nonparametric noise modeling over parametric noise modeling for preserving the TF design of the original function when performing DVR of uncertain scalar fields. For DVR of uncertain data, we adopt a probabilistic view of TFs since it allows incorporating uncertainty in both classification and visual parameter mapping. A probabilistic view is proposed by Drebin \etal~\cite{Drebin88}, where the application of TFs involves two steps: (1) map each sample to a set of material probabilities, and assign each material an RGBA color, and (2) compute the color for each sample as a weighted average of material colors based on material probabilities. Please refer to \SC{sec:dvrQuantile} for more details regarding the probabilistic view. 

\subsection{Multidimensional Transfer Functions}
Traditionally, a TF classifies voxels to optical properties, such as colors and opacity, according to a 1D function of the scalar values. 
The function can be designed either manually, which is a tedious task, or automatically based on attributes of the underlying volume data~\cite{pfister2001transfer}. Since 1D TF classification has limited power in exploring and classifying the embedded features in the data, subsequent studies have considered TFs with multiple dimensions. Multidimensional TFs, as proposed by Kniss \etal~\cite{kniss2001interactive}, have been proven superior to traditional 1D transfer functions due to their ability to isolate complex materials with overlapping intensities. In particular, the gradient magnitude and second-order derivatives are commonly used as additional properties to expand the TF domain~\cite{higuera2004automatic, svakhine2005illustration, ME18}. In this work, we demonstrate the benefits of incorporating gradient magnitude uncertainty, computed analytically within the reconstruction stage into a 2D TF, where a 2D TF is characterized by intensity and gradient magnitude. Broadly speaking, we advocate the extension of a methodology involving the integration of 1D TFs against data uncertainty~\cite{Entezari2016Statistical} to 2D TFs for the improved efficiency and reliability of DVR classifications.

\delete{We investigate material classification in uncertain data with multi-Dimensional TFs. In other words, we re-define multi-dimensional TF classification for uncertain multi-fields, by an analytical derivation of uncertain gradient field, which contrasts with existing methods on table-lookup viewpoint on applying TFs. }

\section{Interpolation of Uncertain Scalar Field for DVR} \label{sec:review}

\subsection{Mathematical Model and the State of the Art}\label{sec:stateOftheArt}
\subsubsection{Interpolation of Parametric Distributions}\label{sec:boxsplineParametric}
We state the mathematical model for our methods and briefly revisit the state of the art in interpolation of uncertain data when the uncertainty is modeled as probability distributions. Given $3$D discrete uncertain scalar data $f(\vv_i)$, the uncertainty can be modeled at each voxel $\vv_i$ by a random variable $X_i$. We assume random variables modeling uncertainty at voxels to be independent. The reconstruction of the random field at an arbitrary position, $\vv$, results in a random variable $X$, which is a linear combination of random variables at positions $\vv_i$'s: $X = \sum_i w_i X_i$, with weights $w_i = \varphi(\vv -\vv_i)$, where $\varphi: \rthree \to \reel$ is the basis function that determines the weights for the neighboring voxels contributing to the interpolated sample $\vv$. When the data are sampled on a Cartesian grid, $\varphi$ is commonly chosen as a tensor product-based trilinear B-spline. The goal is to analytically obtain the probability distribution of $X$ at any arbitrary sampling point $\vv$. 
The probability distribution of a linear combination of independent random variables is the convolution of their individual distributions~\cite{hoggintroduction}. Therefore, the \ta{PDF} of $X$ can be derived from the convolution of the \ta{PDFs} of $X_i$', i.e., $\text{pdf}_{X}(x) = \text{pdf}_{w_1 X_1}(x) *\cdots * \text{pdf}_{w_K X_K}(x)$, 
where pdf$_{X}$ represents the uncertainty at the interpolated point, and pdf$_{w_i X_i}$ denotes the scaled distribution of a random variable $X_i: \text{pdf}_{w_i X_i}(x) = \frac{1}{w_i} \text{pdf}_{X_i}(x / w_i)$, for $1 \leq i \leq K$.

A common choice for modeling uncertainty is the Gaussian distribution. When \ta{PDFs} of $X_i$'s are represented as Gaussian distributions, the \ta{PDF} of the interpolated point $X$ is also a Gaussian distribution whose mean and variance are linearly transformed from the means and variances of $X_i$'s~\cite{feller68}. Liu \etal~\cite{TA:2012:LLBP} modeled \ta{PDFs} of $X_i$ using Gaussian mixture models (GMM). Their framework comprised fitting a GMM to uncertain data $f(\vv_i)$ using expectation maximization~\cite{TA:1977:DLR}, and involved \ta{MC} sampling of estimated GMM for the \ta{PDF} estimation at the interpolation point $\vv$. Sakhaee \etal~\cite{Entezari2016Statistical} modeled \ta{PDFs} of $X_i$'s using compactly supported box splines, e.g.,  uniform distributions, and demonstrated the benefits of a box spline framework over a Gaussian assumption for DVR of uncertain data. 

 
 \delete{Each of these vectors is the shadow of an edge of the N-hypercube adjacent to its origin.
Notationally, we denote the $d$-variate box spline with directions $[\bxi_1, \bxi_2, ..., \bxi_N]$ in $\reel^d$ as $M_{[\bxi_1, \bxi_2, ..., \bxi_N]}(\vx)$. The elementary box splines, $M_{\bxi_n}$, are uniform distributions with one direction specified by $\bxi_n$; the uniform distributions are supported over $\vx = t\bxi_n$ with $t \in [0, 1]$.
The key property of box splines is that the space of box splines is closed under convolution. In other words, the convolution of multiple box splines $M_{\bxi_1} * \cdots * M_{\bxi_N}$ is another box spline $M_{[\bxi_1, \bxi_2, ..., \bxi_N]}(\vx)$ specified by all individual direction vectors.\delete{ I would REMOVE this FIGURE, it takes space, doesn't add to contribution of the paper, and makes the paper totally look like the previous work: For example,~\FG{fig:bivariateBoX} demonstrates the construction of box splines in $\reel^2$ with $1$, $2$, $3$, and $4$ direction vectors. Instead, we can say:} \ellie{We refer the interested readers to~\cite{deBoor2001} for details on construction of box splines.} Box splines are efficiently evaluated via de Boor-H{\"o}llig recurrence~\cite{entezari08}. } 

\subsubsection{Quantile Interpolation of Histograms in 1D}\label{sec:1dQuantileInterpolation}

Even though the box-spline method allows for interpolation of nonparametric distributions in a closed form~\cite{Entezari2016Statistical} (described again in the supplementary material for this paper), it is computationally expensive because of its exponential nature. The exponential time complexity of the convolution-based nonparametric approach is a challenge to handle in the DVR reconstruction stage even when the system has high-performance hardware. In our contribution, we take advantage of the linear time complexity quantile interpolation technique for interpolation of nonparametric distributions. Read~\cite{TA:1999:Read} proposed a quantile interpolation method for 1D interpolation of histograms. To summarize their approach, a histogram or nonparametric density characterizing the \ta{PDF} at each vertex $\vv_i$ is broken into a fixed number of quantiles, and the respective quantiles at each vertex are interpolated to compute a probability distribution at the interpolated point v. Thus, the computational complexity of computing the \ta{PDF} at an interpolated position is \emph{linearly} proportional to the number of quantiles.

The use of quantile interpolation in the context of uncertainty visualization was first advocated by Hollister and Pang~\cite{TA:2013:HP}. The quantile interpolation method was shown to have two desirable qualities in the context of uncertainty visualization. First, quantile interpolation preserves the modality of probability distributions at cell vertices $\vv_i$. Second, the variance of interpolation data is thresholded from below by variances of probability distributions at grid vertices. Moreover, quantile interpolation of histograms has been shown to better capture the shape of the interpolated distribution than the interpolation of parametric or GMM distributions. The quantile interpolation allows for a closed-form solution of the \ta{PDF} at the interpolated position, and hence allows for efficient and accurate \ta{PDF} computations.

Let $\text{pdf}_{X_1}(x)$ and $\text{pdf}_{X_2}(x)$ be the continuous probability distributions \ta{for random variables $X_1$ and $X_2$} at 1D cell vertices $\vv_1$ and $\vv_2$, respectively. The probability distributions $\text{pdf}_{X_1}(x)$ and $\text{pdf}_{X_2}(x)$ can be estimated from noise samples using histograms or kernel density estimation \ta{(KDE)}~\cite{TA:1956:R, TA:1962:P}. Suppose the distributions $\text{pdf}_{X_1}(x)$ and $\text{pdf}_{X_2}(x)$ are broken into $q$ quantiles each, where $qval$ denotes the quantile value. For example, setting $qval=0.5$, $qval=0.25$, and $qval=0.125$ results in median (q=2), quartiles (q=4), and octiles (q=8), respectively.

\ta{Let $Q_{i1}, Q_{i2} \cdots, Q_{iq}$ denote $q$ quantiles ordered by cumulative density function (CDF) with widths $w_{i1}, w_{i2} \cdots, w_{iq}$, respectively, for a random variable $X_i$ associated with grid vertex $\vv_i$ when the quantile value is set to $qval$. The quantile representation for the PDF of each random variable $X_i$ is a piecewise constant function in which each quantile (piece) $j$ assumes a constant probability density, i.e., $Pr(Q_{ij}) = \frac{qval}{w_{ij}}$. We, therefore, present $pdf_{X_i}(x)$ as a tuple $\{Pr(Q_{i1}), \cdots, Pr(Q_{iq})\}$ in its quantile representation, in which each entry of a tuple denotes the probability density over its respective quantile. Let $\alpha$ indicate the spatial distance parameter between 1D cell vertices $\vv_1$ and $\vv_2$. Let $\{Q_1, \cdots Q_q\}$ denote the quantile representation for a linearly interpolated random variable $X = \alpha X_{2} + (1 - \alpha) X_{1}$ when the quantile value is $qval$. Based on the previous work in~\cite{TA:1999:Read} and~\cite{TA:2015:HP}, the probability density for the j'th quantile $Pr(Q_j)$ of the interpolated random variable can be computed as follows: 

  
\begin{align*} Pr(Q_j) &= \frac{Pr(Q_{1j})Pr(Q_{2j})}{(1-\alpha)Pr(Q_{2j}) + \alpha Pr(Q_{1j})}  \\
	 &=  \frac{qval}{\alpha w_{2j} + (1 - \alpha) w_{1j}}\numberthis
	 \label{eq:1dQuantileInterpolation}
  \end{align*} 
where $\alpha \in [0,1]$  and $j \in \{1, 2, \cdots , q\}$. As can be seen from Eq.~\ref{eq:1dQuantileInterpolation}, the quantile width for the j'th quantile (quantile value = $qval$) of a linearly interpolated random variable $X$  is essentially a linear interpolation of j'th quantile widths of random variables at grid vertices. Since the arithmetic operations in Eq.~\ref{eq:1dQuantileInterpolation} are applied quantile-wise for each quantile $j \in \{1, 2, \cdots , q\}$, the computational complexity of the quantile interpolation is linearly proportional to the number of quantiles $q$. As the number of quantiles $q$ approaches infinity, the PDF at an interpolated position converges to a closed-form continuous probability distribution.
}

\subsection{Quantile Interpolation of Nonparametric Distributions in 3D}\label{sec:3dQuantileInterpolation}

Recently, Hollister and Pang~\cite{TA:2015:HP} presented a closed-form solution for a bilinear interpolation of histograms/nonparametric distributions on a 2D grid using the quantile interpolation technique. We extend the derivation for quantile interpolation of histograms or nonparametric distributions to a 3D case using a similar approach, as proposed in~\cite{TA:2015:HP}. \ta{Let $\alpha$, $\beta$, and $\gamma$ denote the spatial distance parameters for vertices $\vv_i$ of a 3D grid cell along three dimensions. Deriving quantile interpolation in the interior of a 3D cell comprises three steps: first, an interpolated PDF along the cell edges can be computed by applying Eq.~\ref{eq:1dQuantileInterpolation} to PDFs at grid vertices with parameter $\alpha$. Second, the interpolated PDFs in the interior of cell faces can be computed by again applying Eq.~\ref{eq:1dQuantileInterpolation} to interpolated PDFs computed in step one with parameter $\beta$ . Finally, the interpolated PDF in the interior of 3D cell can be computed by applying Eq.~\ref{eq:1dQuantileInterpolation} to interpolated PDFs computed in step two with parameter $\gamma$.} For brevity, we represent the probability $Pr(Q_{ij})$ as $Pr_i$ for the j'th quantile of random variable $X_i$. Then the formula for the interpolated PDF of the j'th quantile, i.e., $Pr(Q_j)$, in 3D is as follows:

\begin{equation}
	Pr(Q_j) = \frac{Pr_1Pr_2Pr_3Pr_4Pr_5Pr_6Pr_7Pr_8}{t_1t_2t_3t_4t_5t_6t_7}
	\label{eq:3dQuantileInterpolation}
\end{equation}  
where:
$t_1 = \alpha Pr_1 + (1 - \alpha)Pr_2$, 
$t_2 = \alpha Pr_3 + (1 - \alpha)Pr_4$, \\
$t_3 = \alpha Pr_5 + (1 - \alpha)Pr_6$,
$t_4 = \alpha Pr_7 + (1 - \alpha)Pr_8$, \\
$t_5 = \beta Pr_1 Pr_2/t_1 + (1 - \beta) Pr_3 Pr_4/t_2$,\\
$t_6 = \beta Pr_5 Pr_6/t_3 + (1 - \beta) Pr_7 Pr_8/t_4$, \\
$t_7 = \gamma Pr_1 Pr_2 Pr_3 Pr_4/(t_1 t_2 t_5) + (1 - \gamma) Pr_5 Pr_6 Pr_7 Pr_8/(t_3 t_4 t_6)$, \\
$\alpha \in [0,1]$,  $\beta \in [0,1]$, and $\gamma \in [0,1]$.\\

\ta{Similar to Eq.~\ref{eq:1dQuantileInterpolation}, Eq.~\ref{eq:3dQuantileInterpolation} simplifies to $Pr(Q_j) = \frac{qval}{w_j}$, where $w_j$ denotes the width of the j'th quantile of an interpolated random variable $X$. The width $w_j$ is essentially a trilinear interpolation of widths of j'th quantiles for PDFs of random variables $X_1, \cdots, X_8$ at 3D cell vertices, evaluated with interpolation parameters $\alpha$, $\beta$, and $\gamma$.}  We compute the formula in Eq.~\ref{eq:3dQuantileInterpolation} using the MATLAB Symbolic Math Toolbox. We validate Eq.~\ref{eq:3dQuantileInterpolation} through an experiment on synthetic data. For our experiment, we define histograms representing continuous probability distributions at eight vertices of a 3D cell. We randomly draw $2 \times 10^6$ samples from each histogram and perform KDE on the samples to estimate continuous PDFs, $\text{pdf}_{X_1}(x) \cdots \text{pdf}_{X_8}(x)$, from the samples. A fixed number of quantiles are then computed for estimated PDFs at each of the eight vertices with the quantile value $qval = 0.001$. Setting the quantile value to $0.001$ results in $q=1000$ quantiles at each cell vertex. The PDF at the interpolated position is then computed in closed form using Eq.~\ref{eq:3dQuantileInterpolation}. \FG{fig:quantileInterpolationTechnique}(a) and (f) visualize the PDF computed at an interpolated position using MC sampling and our analytic derivation, respectively. Having the same shape for both distributions confirms the correctness of Eq.~\ref{eq:3dQuantileInterpolation}.

We compare an interpolated PDF shape computed using quantile interpolation with two GMM models, i.e., the GMM (ordered)~\cite{TA:2013:HP} and the GMM (MC)~\cite{TA:2012:LLBP}. The GMM (ordered) model is a variant of the GMM (MC) model in that Gaussians in each GMM are ranked based on sorted Gaussian means. The Gaussians at each grid vertex with the same rank are then interpolated. The PDF interpolation proposed in~\cite{TA:2012:LLBP} resorts to MC sampling because of its exponential computational complexity. \FG{fig:quantileInterpolationTechnique}(b) shows the result of the interpolated PDF using the GMM (ordered) model, which emphasizes the peaks of the interpolated PDF seen in \FG{fig:quantileInterpolationTechnique}(a) and is smoother. \FG{fig:quantileInterpolationTechnique}(c) shows the result for the GMM (MC) model, in which a lack of ordering among Gaussians produces a PDF shape significantly different from the quantile interpolation or GMM (ordered) models. \tma{We assume four Gaussians for the GMM results since they require storage comparable to the quantile interpolation with eight quantiles (for more details, please see \SC{sec:exp}).}

\ta{\FG{fig:quantileInterpolationTechnique}(d) and (e) visualize quantile interpolation PDF results when the PDF at each grid vertex is summarized using coarse quantile representation. The interpolated PDF with four quantiles in \FG{fig:quantileInterpolationTechnique}(d) appears similar to a uniform distribution without providing any useful insight regarding which data values have a higher probability of occurrence, whereas the interpolated PDF in \FG{fig:quantileInterpolationTechnique}(e) captures the shape of the interpolated PDF reasonably well for an eight-quantile PDF representation. \FG{fig:quantileInterpolationTechnique}(g-i) show results similar to those in \FG{fig:quantileInterpolationTechnique}(d-f), respectively, except that the number of noise samples at the grid vertices is reduced from $2 \time 10^6$ to $2 \time 10^4$. For the reduced number of noise samples, the quantile interpolation results can fluctuate significantly for a relatively high number of quantiles, e.g., \FG{fig:quantileInterpolationTechnique}(f), because of poor KDE caused by reduced sample size.}

\begin{figure} [!ht] 
\centering
\includegraphics[width = 3.2in]{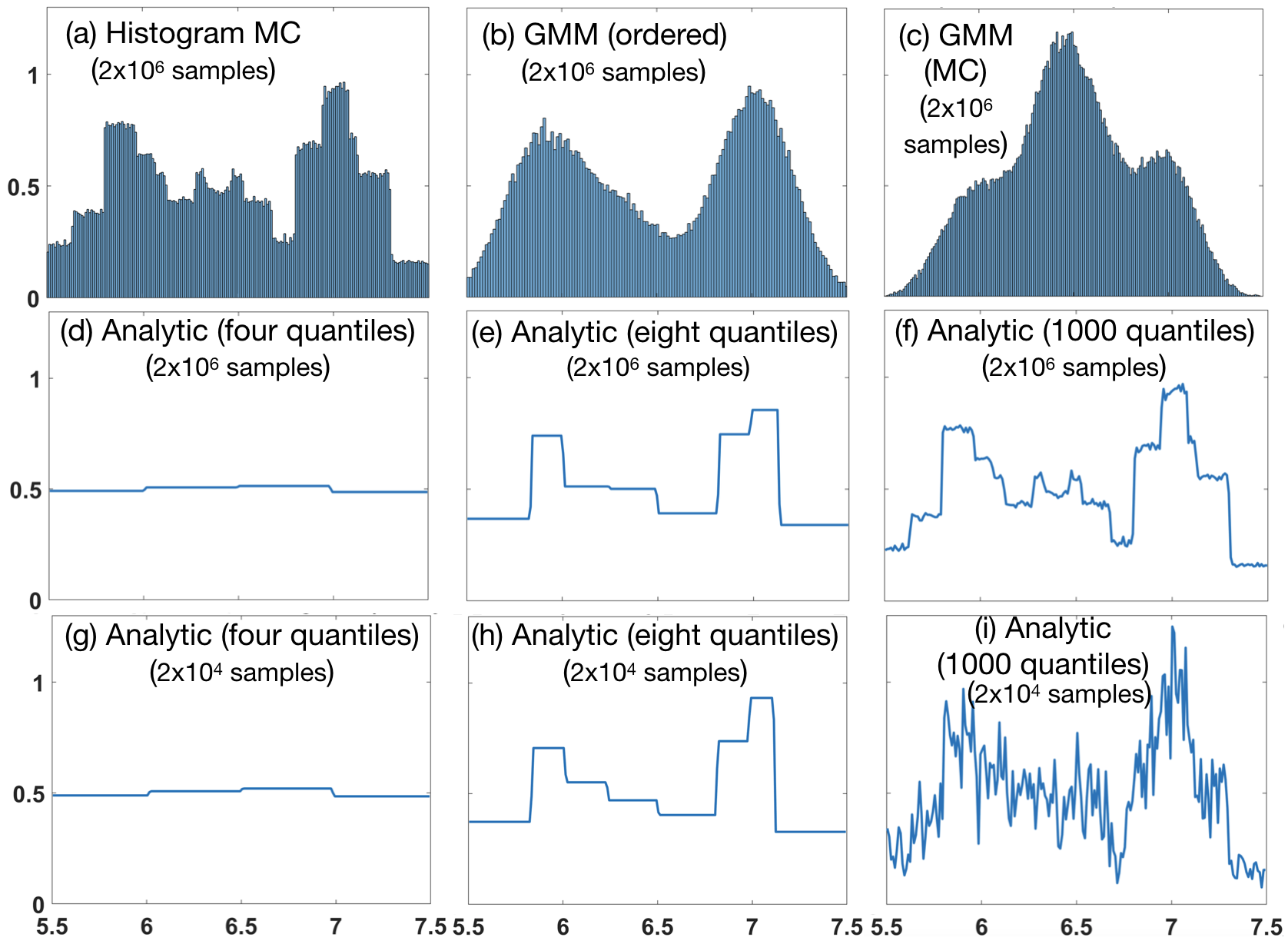}
	\caption{A probability distribution at a 3D interpolated position computed using \ta{MC} sampling (image (a)) vs. our analytic formula (image (f)) for quantile interpolation (Eq.~\ref{eq:3dQuantileInterpolation}). \ta{Images (b) and (c) visualize interpolated PDFs for GMM noise models. Images (d) and (e) visualize interpolated PDFs for coarse quantile representations. Images (g-i) show plots similar to images (d-f), respectively, for reduced sample size.}}
	\label{fig:quantileInterpolationTechnique}
\end{figure} 



\subsection{DVR with Quantile Interpolation}\label{sec:dvrQuantile}

We describe a three-step approach for DVR of uncertain data when data uncertainty is characterized by histograms or nonparametric distributions. In the first step, we preprocess uncertain data. At each grid position, we estimate histogram or nonparametric density from noise samples. We then partition the continuous distribution $\text{pdf}_{X_i}(x)$ at each grid position $\vv_i$ into $q$ quantiles \ta{based on a user-set quantile value $qval$ and compute a quantile representation of $\text{pdf}_{X_i}(x)$}. \ta{We denote quantile representation of $\text{pdf}_{X_i}(x)$ as a tuple $\{Pr(Q_{i1}), \cdots, Pr(Q_{iq})\}$ (see \SC{sec:1dQuantileInterpolation} for details). The quantile representations for  PDFs are provided as inputs to the DVR framework.}

In the second step, for each sample along a viewing ray of the DVR framework, we look up quantile-based PDFs at the neighboring eight vertices in a fragment shader. \ta{We apply Eq.~\ref{eq:3dQuantileInterpolation} to neighboring vertex quantile densities and compute, in closed form, a quantile representation $\{Pr(Q_1), \cdots, Pr(Q_q)\}$ for a continuous distribution $\text{pdf}_{X} (x)$ of the interpolated random variable $X$.} In the third step, the opacity and color for an interpolated random variable $X$ can be computed by applying the uncertainty integration framework proposed by Sakhaee and Entezari~\cite{Entezari2016Statistical} as follows: 

\begin{equation}
	E(\text{TF}(X)) = \int \text{TF}(x) \text{pdf}_{X} (x) dx
	\label{eq:nonparametricIntegral}
\end{equation}
where TF$(x)$ is the color and opacity sampled at intensity $x$ in the TF domain, and $E(\text{TF}(X)$) represents the expected value of the classified color and opacity. \ta{The piecewise constant quantile representation for interpolated $\text{pdf}_{X} (x)$, i.e., a tuple $t = \{Pr(Q_1)=\frac{qval}{w_1}, \cdots, Pr(Q_q)=\frac{qval}{w_q}\}$, may be integrated with a TF in different ways to produce meaningful DVR visualizations.  We propose two integration schemes, namely \emph{quantile range} and \emph{quantile mean}.

\paragraph{Quantile range technique:} In this method, we integrate all intensities contained in a quantile with a TF.  Suppose [a,b] denote the domain of interpolated random variable $X$. By substituting $\text{pdf}_{X} (x)$ as a quantile-based piecewise constant density function represented by the tuple $t$ in Eq. \ref{eq:nonparametricIntegral}, we get the following formula for the expected fragment color:

\begin{align*}
	E(\text{TF}(X)) &= \int_{a}^{a+w_1} \text{TF}(x) Pr(Q_1)dx + \cdots + \int_{b-w_q}^{b} \text{TF}(x) Pr(Q_q) dx\\
			       &=  \frac{qval}{w_1}\int_{a}^{a+w_1} \text{TF}(x)dx + \cdots + \frac{qval}{w_q}\int_{b-w_q}^{b} \text{TF}(x) dx \numberthis
	\label{eq:quantileRangeIntegral}
\end{align*}
In summary, the expected fragment color can be computed by averaging the TFs for each quantile and performing a weighted sum of average colors computed for each quantile, where the weight is equal to $qval$.
   
\paragraph{Quantile mean technique:} In this method, we integrate quantile mean intensities with the TF. Let $m_1, \cdots m_q$ denote the quantile means for quantiles $Q_1 \cdots Q_q$, respectively, of interpolated random variable $X$. Then the discrete probability density for the mean of j'th quantile is $Pr(x = m_j) = \frac{p_j}{p}$, in which $p_j = \frac{qval}{w_j}$, and $p$ is a normalization constant, i.e., $\frac{qval}{w_1} + \cdots +\frac{qval}{w_q}$. The discretized version of Eq. \ref{eq:nonparametricIntegral} for the quantile mean technique results in the following formula for the expected fragment color:

\begin{equation}
	E(\text{TF}(X)) = \sum_{j=1}^{q} \text{TF}(x = m_j) \text{Pr} (x = m_j)
	\label{eq:quantileMeanIntegral}
\end{equation}

}

\paragraph{Memory and Time Complexity:} \label{sec:memoryReq}We comment on the memory and time complexity of our approach for DVR with nonparametric statistics. The preprocessing step is computationally expensive. It involves computing KDE at each grid vertex from noise samples~\cite{TA:1956:R, TA:1962:P} and reducing continuous distributions to \ta{their quantile representations}. Let \ta{$L \times M \times N$} denote the grid size, and $q$ denote the number of quantiles at each grid vertex. The memory consumed by input data for our nonparametric DVR framework is, therefore, \ta{$L \times M \times N \times (q+1)$}, in which $q+1$ represents the number of boundaries defining quantiles at each grid vertex. Thus, the memory requirement for our nonparametric framework increases linearly with the number of quantiles. The fragment shader for our DVR framework computes the distribution at viewing-ray samples through quantile interpolation of neighboring \ta{quantile representations of PDFs} with $q$ quantiles (Eq.~\ref{eq:3dQuantileInterpolation}). The \ta{computational complexity of} the quantile interpolation is linearly proportional to the number of quantiles because of the \ta{quantile-wise} arithmetic of probability distributions in Eq.~\ref{eq:3dQuantileInterpolation}. Hence, the computational complexity of our DVR framework increases linearly with the number of quantiles.



%
\ta{

\subsection{Quartile View} \label{sec:quartileView}

The quantile interpolation is an example of order statistics, where quantiles are ranked based on CDF for a random variable. Specifically, the j'th quantile (ranked by CDF) of input probability distribution $\text{pdf}_{X_i}(x)$ corresponds to the j'th quantile of a distribution computed using quantile interpolation (see Eq.~\ref{eq:3dQuantileInterpolation}). We take advantage of the order statistics exhibited by quantile interpolation to derive a box-plot-like view for the input data, which we refer to as the quartile view. Specifically, we visualize populations corresponding to three quantiles, i.e., first 25\% (lower quartile), middle 50\%, and last 25\% (upper quartile), in the quartile view. We compute the expected color for each of the three populations with our quantile range method (Sec.~\ref{sec:dvrQuantile}). The middle 50\% population represents a visualization corresponding to the interquartile range (IQR) for the input data, which is considered as a robust range in order statistics. Also, the quartile view can help users understand uncertainty in input data by examining commonalities and differences among visualizations for the lower, middle 50\%, and upper quartiles. Our results in \FG{fig:tangleUncertainty} and \FG{fig:kaustUncertainty} illustrate the quartile views.
}

\section{2D TF for Classification of Uncertain Data} \label{sec:gradient}
\subsection{Interpolation of Uncertain Gradient Field} \label{sec:gradientInterp}
The reconstruction of the gradient field from the scalar field is an essential process in volume rendering~\cite{TA:2003:MZBQ}. The gradient information can be employed in both multidimensional TFs (for material classification)~\cite{Drebin88} and various local illumination techniques~\cite{Hadwiger06}.
The gradient is mathematically the first-order derivative of the scalar field $f(\vv)$, $\grad f(\vv) = \transp{(\frac{\partial f(\vv)}{\partial x}, \frac{\partial f(\vv)}{\partial y}, \frac{\partial f(\vv)}{\partial z})}$, and points in the direction of the steepest ascent. The reconstruction of the gradient field at an arbitrary position $\vv$ in an uncertain scalar field results in a trivariate random vector $\vect{Y} = \transp{(Y_x, Y_y, Y_z)}$, where $Y_x, Y_y, \text{and }Y_z$ are random variables denoting the uncertainties of a partial derivative along directions $x, y, \text{and }z$, respectively. The uncertain gradient $\vect{Y}$ can be obtained from linear combinations of random variables at neighboring voxels $\vv_i$'s of $\vv$:


\begin{equation} \label{eq:computeGradient}
	\left[{\begin{array}{cc}
   Y_x \\ Y_y \\ Y_z 
   \end{array} } \right] = \sum_i \left[{\begin{array}{cc}
   \alpha_i \\ \beta_i \\ \gamma_i 
   \end{array} } \right] X_i \ \  \text{with weights} \ \  \left[{\begin{array}{cc}
   \alpha_i \\ \beta_i \\ \gamma_i 
   \end{array} } \right] = \rho(\vv -\vv_i)
\end{equation}
where $\rho: \rthree \to \rthree$ is the basis function that determines the amount of the contribution of each neighboring \ellie{random variable} $X_i$ to gradient $\vect{Y}$ at $X$. 
The basis function $\rho$ can be represented as a derivative reconstruction filter obtained by convolving a continuous interpolation filter with a digital derivative filter~\cite{moeller97}. 
In the context of volume rendering, the most common choice is the combination of linear interpolation and central difference. For voxels that fall out of the support of $\rho$, zero weights are assigned. As a result, only $M$ neighbors of $X$ are involved in the gradient estimation, i.e., $1 \leq i \leq M$.

As shown in Eq.~\ref{eq:computeGradient}, the uncertain gradient estimation at arbitrary point $\vv$ is also obtained by linear combinations of random variables $X_i$'s of the uncertain scalar field. Therefore, the distribution of the gradient at $\vv$ can also be derived analytically using the spline framework. For example, when data uncertainties are modeled as (scaled) uniform distributions, the distribution of each partial derivative is also a box spline whose direction vectors are the interpolation weights. In this setup, the joint distribution of the random vector $\vect{Y}$, i.e., the uncertain gradient, is analytically derived as a trivariate box spline whose directions are obtained by the weights in Eq.~\ref{eq:computeGradient}, i.e., the $i$-{th} direction vector is $\transp{[\alpha_i, \beta_i, \gamma_i]}$.
\delete{For example, when data uncertainties are modeled as uniform distributions pdf$_{X_i} = \textit{U}(0, 1)$, the distribution of each partial derivative is also a box spline with all the weights as the direction vectors, e.g., pdf$_{Y_x}(x) = M_{[\alpha_1, \dots, \alpha_M]}(x)$. In this setup, the joint distribution of the random vector $\vect{Y}$, i.e., the uncertain gradient, is analytically depicted by a trivariate box spline $M_{[\bxi_1, \cdots, \bxi_M]}(\vx)$ whose directions $[\bxi_1, \cdots, \bxi_M]$ are obtained by the weights in \EQ{eq:computeGradient}: $\bxi_i = \transp{[\alpha_i, \beta_i, \gamma_i]}$.}

\begin{figure*} [!ht] 
\centering
\includegraphics[width = 5.8in]{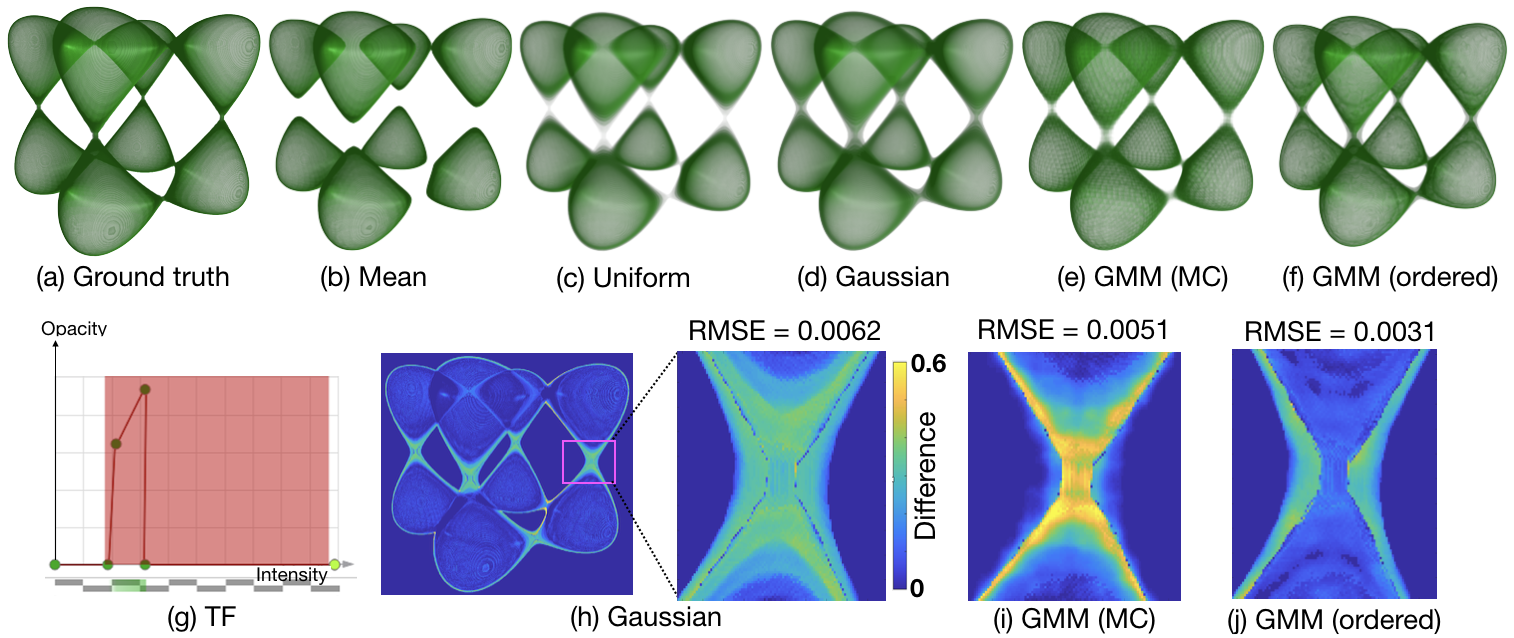} \\
\includegraphics[width = 5.8in]{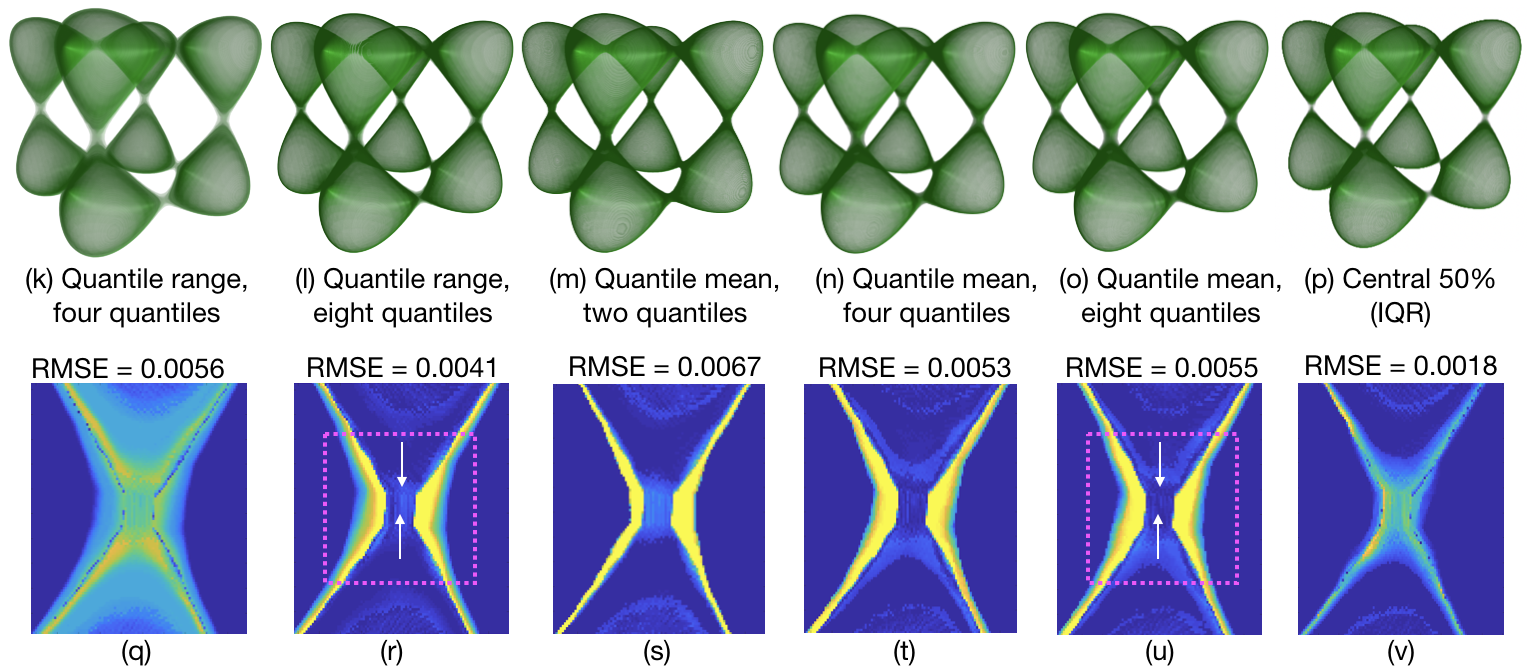}
\caption{\ta{Nonparametric (third and fourth rows) vs. parametric statistics (first and second rows) for visualizations of the tangle function: The ground truth volume visualized in (a) is mixed with noise to generate an ensemble representing uncertain data. The DVR visualizations for various statistical models in the first and third row are rendered with the same TF (image (g)). The second and the fourth rows quantify and visualize the differences with respect to the ground truth for their corresponding DVR images in the first and third rows, respectively. In images (r) and (u), the white arrows illustrate the positions of high reconstruction accuracy and the dotted pink boxes enclose the positions that illustrate error bands.}}
	\label{fig:tangle}
\end{figure*} 

\subsection{2D TF Integration with Interpolated Gradient Field } \label{sec:ProbTF}
Unlike the gradient, it is challenging to analytically derive the PDF of the gradient magnitude given that random variables $Y_x, Y_y, \text{and }Y_z$ are correlated, even when these random variables are modeled as parametric normal distributions~\cite{MP92}. Therefore, the directional derivative along the mean gradient direction has been commonly used as an approximation in uncertainty visualization~\cite{TA:2013:PMW}. Mathematically, the gradient magnitude $\norm{\vect{Y}} \approx \frac{\mu_{\vect{Y}}}{\norm{\mu_{\vect{Y}}}} \vect{Y}$, where $\mu_{\vect{Y}}$ is the mean gradient, which can be obtained by substituting $X_i$'s with their mean values in \EQ{eq:computeGradient}. Since the directional derivative is also linear with respect to the original scalar field, we can formulate the interpolation of intensity $X$ and gradient magnitude $\norm{\vect{Y}}$ as linear combinations of $M$ neighboring voxels of $\vv$,\ta{ i.e., $X =  \sum_{i=1}^{M} w_iX_i$ and $\norm{\vect{Y}}  =  \sum_{i=1}^{M} u_iX_i$}. The joint \ta{PDF} of the bivariate random variable $\vect{Z} = \transp{[X, \norm{\vect{Y}}]}$ is exactly described by a bivariate box spline with parameters $[\transp{[w_1, u_1]}, \cdots, \transp{[w_M, u_M]}]$. Similar to \EQ{eq:nonparametricIntegral}, we compute the color at position $\vv$ by integrating the 2D TF with the joint \ta{PDF} of intensity and gradient magnitude $\vect{Z}$, i.e., $E(\text{TF}(\vect{Z})) = \int \int \text{TF}(x, y) \text{pdf}_{\vect{Z}} (x, y) dx dy$.

\delete{$M_{[\bxi_1, \cdots, \bxi_M]}(x, y)$ whose directions $[\bxi_1, \cdots, \bxi_M]$ are obtained by the weights in the linear system: $\bxi_i = \transp{[w_i, u_i]$. }}
\delete{Examples of the construction of the joint distributions of two random variables are shown in \FG{fig:bivariateBoX}. } 


\section{Results and Discussions} \label{sec:exp}

We demonstrate the effectiveness of our nonparametric noise modeling (\SC{sec:dvrQuantile}) for DVR on two kinds of synthetic and real datasets: ensembles of volumetric datasets (\FG{fig:teaser}a, \FG{fig:tangle}, \ta{and \FG{fig:kaustUncertainVelocity}}) and downsampled versions of high-resolution datasets (\FG{fig:teaser}b, \FG{fig:nestedSpheres}, and \FG{fig:obelix}). Next, we demonstrate improved DVR classifications with $2$D TFs  (\FG{fig:tooth}, \FG{fig:fiber1}), where underlying uncertain data are represented as probability distributions (\SC{sec:gradient}) as opposed to mean statistics. 
 All volume renderings are performed on a machine with \ta{Nvidia GPU Quadro P6000}, with \ta{$24$} GB memory. We integrate the fragment shaders for our statistical frameworks into the Voreen volume rendering engine (\url{http://voreen.uni-muenster.de}) for DVR of uncertain data.

\subsection{Statistical Rendering: Nonparametric vs. mean/parametric statistics}


\paragraph{Ensemble Datasets:} In \FG{fig:tangle}, we perform \ta{qualitative and quantitative assessment of DVR reconstruction accuracy for different noise models on a synthetic tangle~\cite{knoll2009fast} dataset.} \FG{fig:tangle}(a) visualizes the ground truth tangle function for a fixed TF design \ta{(\FG{fig:tangle}(g))}. The same TF is used for all visualizations in \FG{fig:tangle}. The ground truth volume is mixed with noise to generate \ta{an ensemble of $50$ members} representing uncertain data. \ta{Specifically, we inject noise samples randomly drawn from a bimodal probability distribution, in which the mode with 80\% probability concentration is centered around the ground truth, and the mode with 20\% probability concentration (representing outliers) is centered far away from the ground truth. The injected noise, thus, has a shape similar to the one-tailed asymmetric distribution.} 

\ta{\FG{fig:tangle}(b) and \FG{fig:tangle}(c) visualize the results for the mean-field and uniform noise models, respectively. The presence of the outliers in the noise samples shifts the sample mean at each grid vertex substantially, hence breaking the regions connecting the blobs of the tangle function. The reconstruction with the uniform model still shows \ta{improved topological recovery} compared to the mean-field. The visualizations using Gaussian, GMM (MC), and GMM (ordered) in \FG{fig:tangle}(d-f), respectively, and our proposed nonparametric models in \FG{fig:tangle}(k-p) show further reconstruction improvements.} 

\ta{In the second and fourth rows of \FG{fig:tangle}, we perform quantitative analysis of the reconstruction accuracy for Gaussian, GMM (MC), GMM (ordered), and nonparametric statistical models. Specifically, at each pixel, we compute an absolute difference between the mean of the RGB values for a DVR image specific to a noise model and the DVR of the ground truth (\FG{fig:tangle}(a)). We then visualize computed differences using a blue-yellow diverging color map, in which yellow and blue indicate relatively high and low difference regions, respectively. In \FG{fig:tangle}, we also report the root mean squared error (RMSE) for each noise model. Note that the difference images for the mean and uniform noise models are not shown in \FG{fig:tangle} because of their relatively high RMSE values, i.e., $0.0245$ and $0.0200$, respectively. }

\ta{We assume four Gaussians for modeling uncertainty with GMMs since they consume a memory comparable to the quantile interpolation with eight quantiles (for details, see the next paragraph). In the case of the GMM (MC) model, we draw MC samples from a GMM until no significant variations are observed in the visualizations. The same process is followed for all the datasets. In \FG{fig:tangle}(e), we drew $200$ MC samples per grid vertex. The difference images for the GMM (ordered) (\FG{fig:tangle}(j)) and our nonparametric approach (\FG{fig:tangle}(r-u)) show relatively high reconstruction accuracy in the interior of the dataset, as indicated by the white arrows in \FG{fig:tangle}(r) and (u), when compared to the difference images for the Gaussian (\FG{fig:tangle}(h)) and GMM (MC) (\FG{fig:tangle}(i)) noise models. Moreover, the error pattern is systematic and refined in the case of our quantile range and quantile mean techniques with eight quantiles. Specifically, the error bands are observed (enclosed within the pink dotted boxes in \FG{fig:tangle}(r) and (u)) with a relatively high error near the boundary between the dataset and background with the error gradually decreasing as we move farther away from the dataset. The RMSE for our nonparametric approach with eight quantiles is smaller than the ones for mean, uniform, and Gaussian statistics. Also, increasing the number of quantiles for our quantile interpolation techniques decreases the RMSE for tangle visualizations when the number of quantiles is increased from two to four, e.g., \FG{fig:tangle}(q-r). \FG{fig:tangle}(v) visualizes a difference image for the expected visualization in \FG{fig:tangle}(p) corresponding to the middle 50\% quantile (IQR), ordered by CDF, for the random variable at each grid vertex. The RMSE for the IQR visualization is the lowest among all visualizations in \FG{fig:tangle}.}

\ta{The uniform and Gaussian noise assumptions for the visualizations in \FG{fig:tangle}(c) and (d), respectively, require storing the mean and width/variance of uncertain values, and hence consume twice the memory needed for the mean-field approach in \FG{fig:tangle}(b). The visualizations in \FG{fig:tangle}(e) and (f) with four Gaussians per GMM consume $12$ times the memory needed for the mean-field (see~\cite{TA:2012:LLBP}), as each Gaussian requires storing its mean, variance, and weight. Our proposed nonparametric models consume memory linearly proportional to the number of quantiles (\SC{sec:memoryReq}). For example, the quantile representations for the results shown in \FG{fig:tangle}(l) and (o) consume nine times the memory needed for the mean-field. The IQR visualization \FG{fig:tangle}(p) interpolates only a single quantile (middle 50\%), and, thus, requires only twice the memory needed for the mean-field.}



\begin{figure} [!ht] 
\centering
\includegraphics[width = 3.5in]{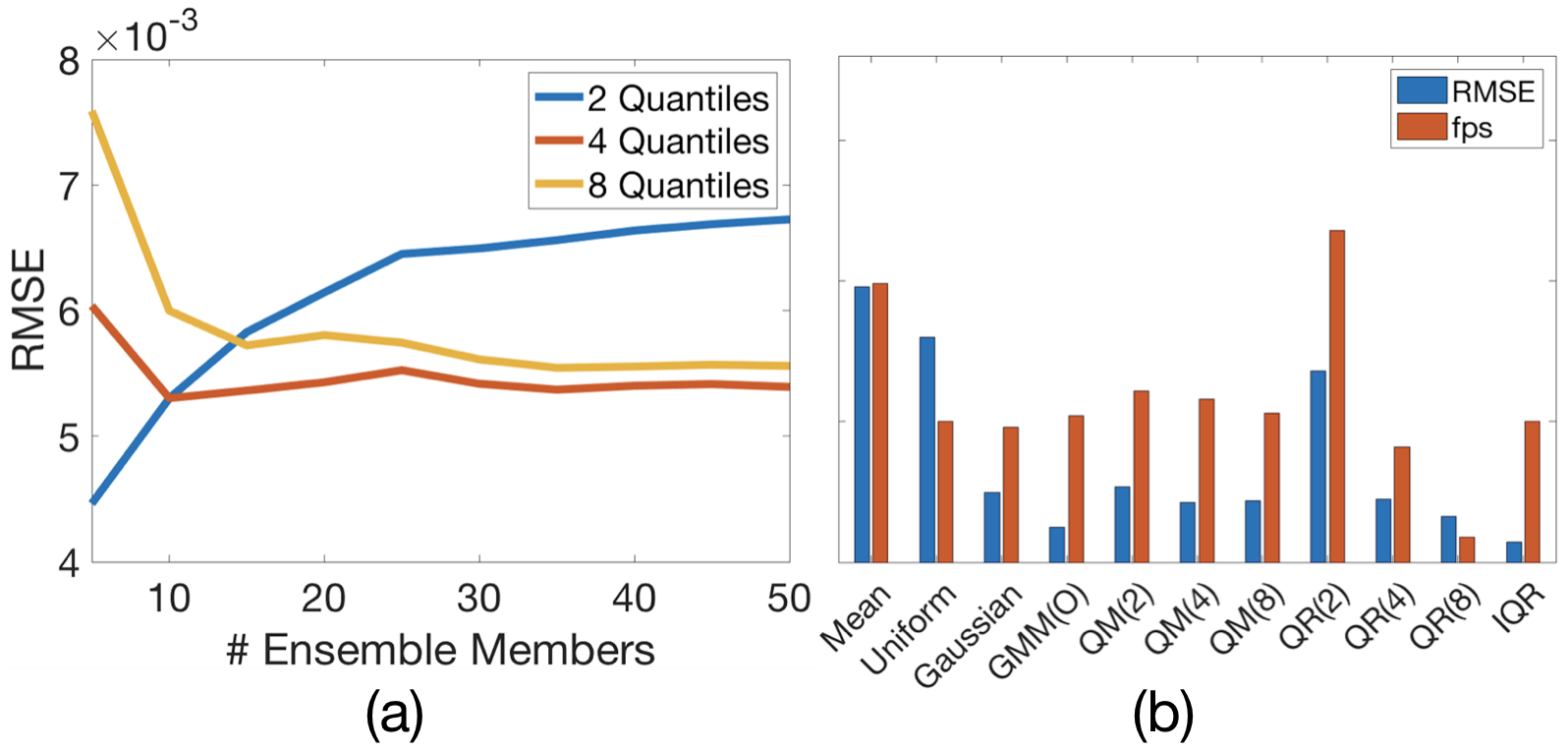}
	\caption{\ta{(a) Effects of number of quantiles and sample size on reconstruction accuracy. (b) The relative RMSE and frame rate (fps) for different noise models. In image (b), GMM (O), QM, and QR denote GMM (ordered), quantile mean, and quantile range noise models, respectively.}}
	\label{fig:tangleQuantileAndSampleNumberEffect}
\end{figure} 

\ta{In \FG{fig:tangleQuantileAndSampleNumberEffect}(a), we analyze the effects of the number of quantiles in our nonparametric framework and the number of ensemble members on the reconstruction accuracy of the tangle dataset. As observed in \FG{fig:quantileInterpolationTechnique}, analytically computed quantile interpolation can be quite unreliable when the average number of samples per quantile is relatively low (\FG{fig:quantileInterpolationTechnique}(i)). We observe a similar trend in \FG{fig:tangleQuantileAndSampleNumberEffect}(a) for the tangle dataset. Analysis with a small sample size, e.g., five ensemble members, results in poor \ta{KDE} at grid vertices, hence resulting in a RMSE for the four/eight quantile case greater than the one for the two quantile case. In contrast, the RMSE for the four/eight quantile case becomes smaller than the RMSE for the two quantile case and becomes more stable, as the number of ensemble members increases to $50$. 

\FG{fig:tangleQuantileAndSampleNumberEffect}(b) visualizes the relative RMSE and frame rate for various noise models. The mean-field is rendered at a frame rate of $10$ fps. We measured a rendering performance of the visualizations of the tangle dataset (\ta{$64 \times 64 \times 64$}) using Nvidia's Frameview tool. Our proposed quantile mean method achieves frame rates comparable to parametric noise models for eight quantile representations. Note that the frame rate is not shown for the GMM (MC) model in \FG{fig:tangleQuantileAndSampleNumberEffect}(b), as the GMM (MC) noise model~\cite{TA:2012:LLBP} uses screen space integration of MC samples for producing static images. \FG{fig:tangleUncertainty} visualizes a quartile view (\SC{sec:quartileView}) for the tangle dataset. The magenta boxes in \FG{fig:tangleUncertainty} highlight the positions that exhibit reconstruction variability across the three quartile populations, hence indicating data uncertainty in those regions.}

\begin{figure} [!ht] 
\centering
\includegraphics[width = 3in]{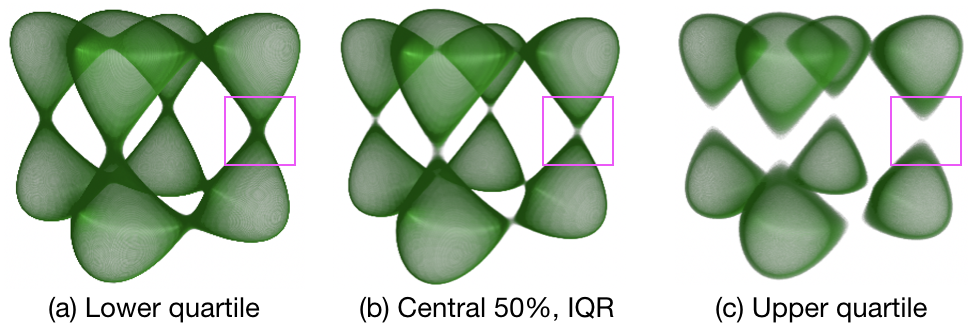}
	\caption{\ta{The quartile view for the uncertain tangle dataset. The pink boxes mark positions that exhibit reconstruction variations across three populations.}}
	\label{fig:tangleUncertainty}
\end{figure}



\FG{fig:teaser}(a) visualizes the synthetic teardrop function~\cite{knoll2009fast} for an experiment similar to the tangle function. \ta{For the teardrop dataset, we again analyze an ensemble of $50$ members.} In \FG{fig:teaser}(a), the reconstruction in the case of the nonparametric density assumption is superior compared to the parametric density assumptions with ground truth as a reference.

\begin{figure} [!ht] 
\centering
\includegraphics[width = 2.8in]{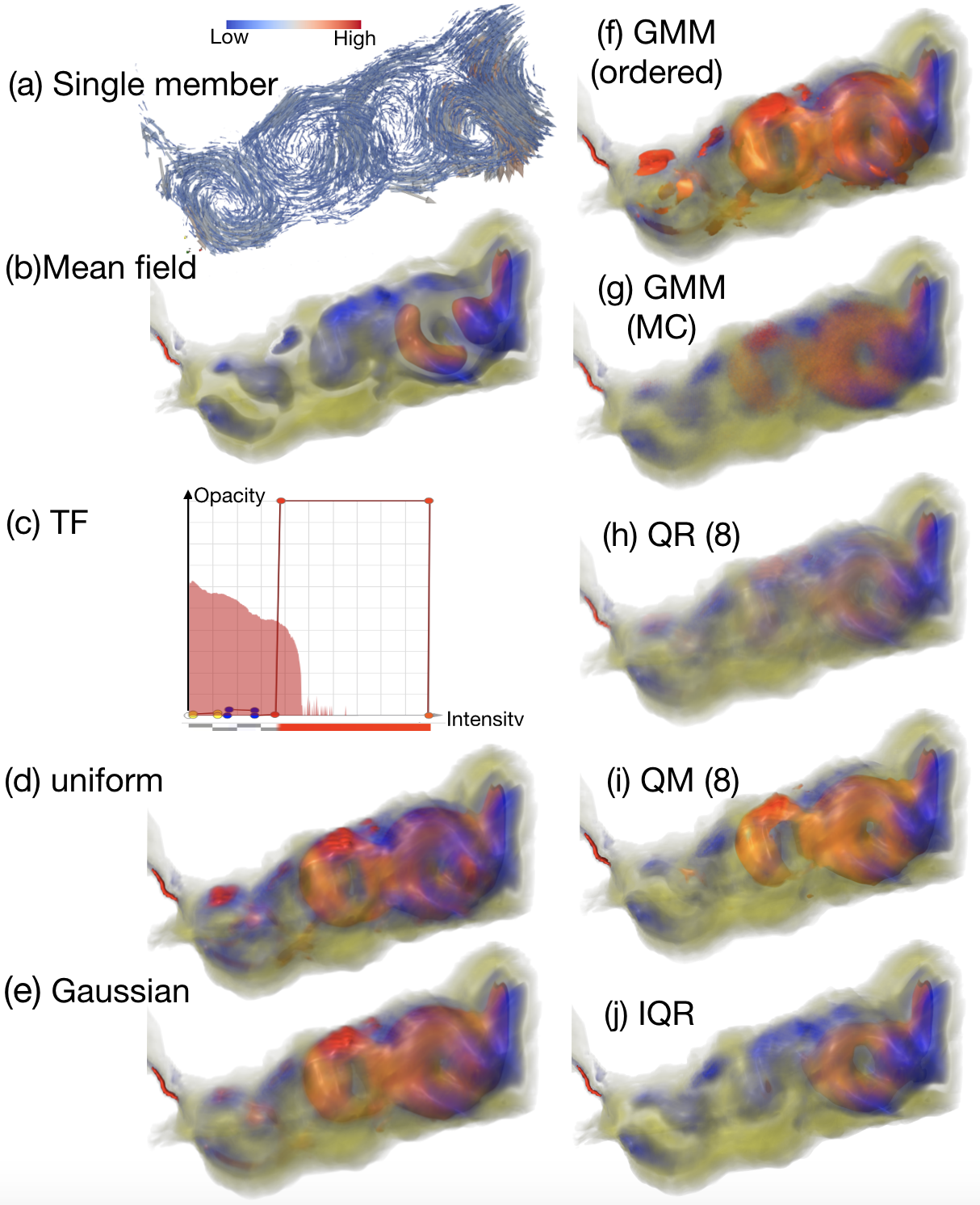}
\caption{\ta{Visualizations of an uncertain velocity magnitude field for the Red Sea eddy simulations: (a) arrow glyph visualization of a velocity vector field for a single member colored by magnitude, (b, d-j) DVR using various statistical models with the TF shown in image (c). The red, blue, and yellow in the TF indicate relatively high-, moderate-, and low-velocity magnitudes. QR(8) and QM(8) denote our proposed quantile range and mean techniques, respectively, with eight quantiles.}}
	\label{fig:kaustUncertainVelocity}
\end{figure} 

We perform an experiment similar to the ones for the tangle and teardrop functions on a real dataset. We analyze the Red Sea eddy simulation ensemble \ta{comprising $20$ members} made available at the IEEE SciVis Contest 2020 (\url{https://kaust-vislab.github.io/SciVis2020/}). Each member of the ensemble dataset is generated based on the MIT ocean general circulation model (MITgcm) and the Data Research Testbed (DART)~\cite{TA:2013:HHGA} with varying initial conditions. The ensembles are sampled for $60$ time steps on a grid with resolution \ta{$500 \times 500 \times 50$} to represent a time-varying 3D flow~\cite{TA:2020:STZL}.

\FG{fig:kaustUncertainVelocity} shows the DVR of the uncertain velocity magnitude field for an ensemble (time step = $40$) over a portion of the domain using the mean-field, parametric, and our nonparametric statistical frameworks. \ta{\FG{fig:kaustUncertainVelocity}(a) visualizes a velocity vector field using arrow glyphs colored by magnitude for a single ensemble member. High-velocity magnitude generally is observed on a vortex rim.} \FG{fig:kaustUncertainVelocity}(b) visualizes the result for the mean-field. We set a TF \ta{(\FG{fig:kaustUncertainVelocity}(c))} for the mean-field visualization, such that relatively high-, moderate-, and low-velocity magnitudes are assigned red, blue, and yellow, respectively. The opacities are set to recover eddy-like structures from the dataset. \ta{The same TF is then used for all visualization in \FG{fig:kaustUncertainVelocity}. The results for all noise models in~\FG{fig:kaustUncertainVelocity} look significantly different from the mean-field visualization. \FG{fig:kaustUncertainty} visualizes a quartile view for the Red Sea dataset. The dotted black boxes in \FG{fig:kaustUncertainty} highlight the positions where the variability or uncertainty of eddy presence is prominent across populations belonging to the lower, central 50\%, and upper quartiles. In contrast, the solid black boxes highlight the positions where eddy presence is consistently observed across the three populations.}



\begin{figure} [!ht] 
\centering
\includegraphics[width = 3.5in]{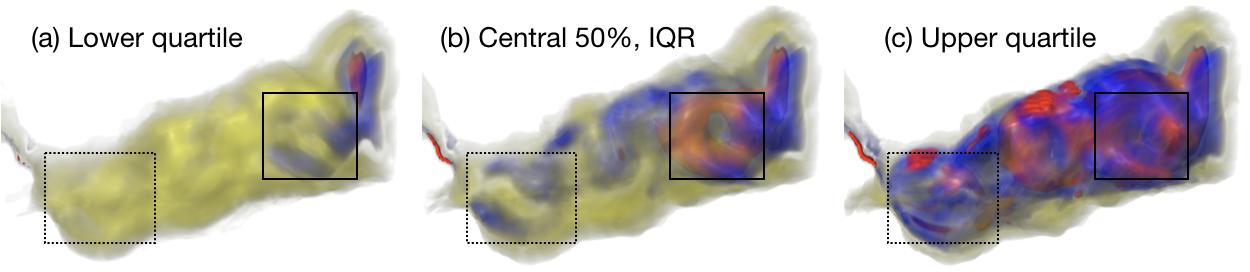}
\caption{\ta{The quartile view for the Red Sea eddy dataset. The solid and dotted boxes indicate the positions with relatively high and low confidence, respectively, regarding the eddy presence.}}
	\label{fig:kaustUncertainty}
\end{figure} 



\begin{figure} [!ht] 
\centering
\includegraphics[width = 3.5in]{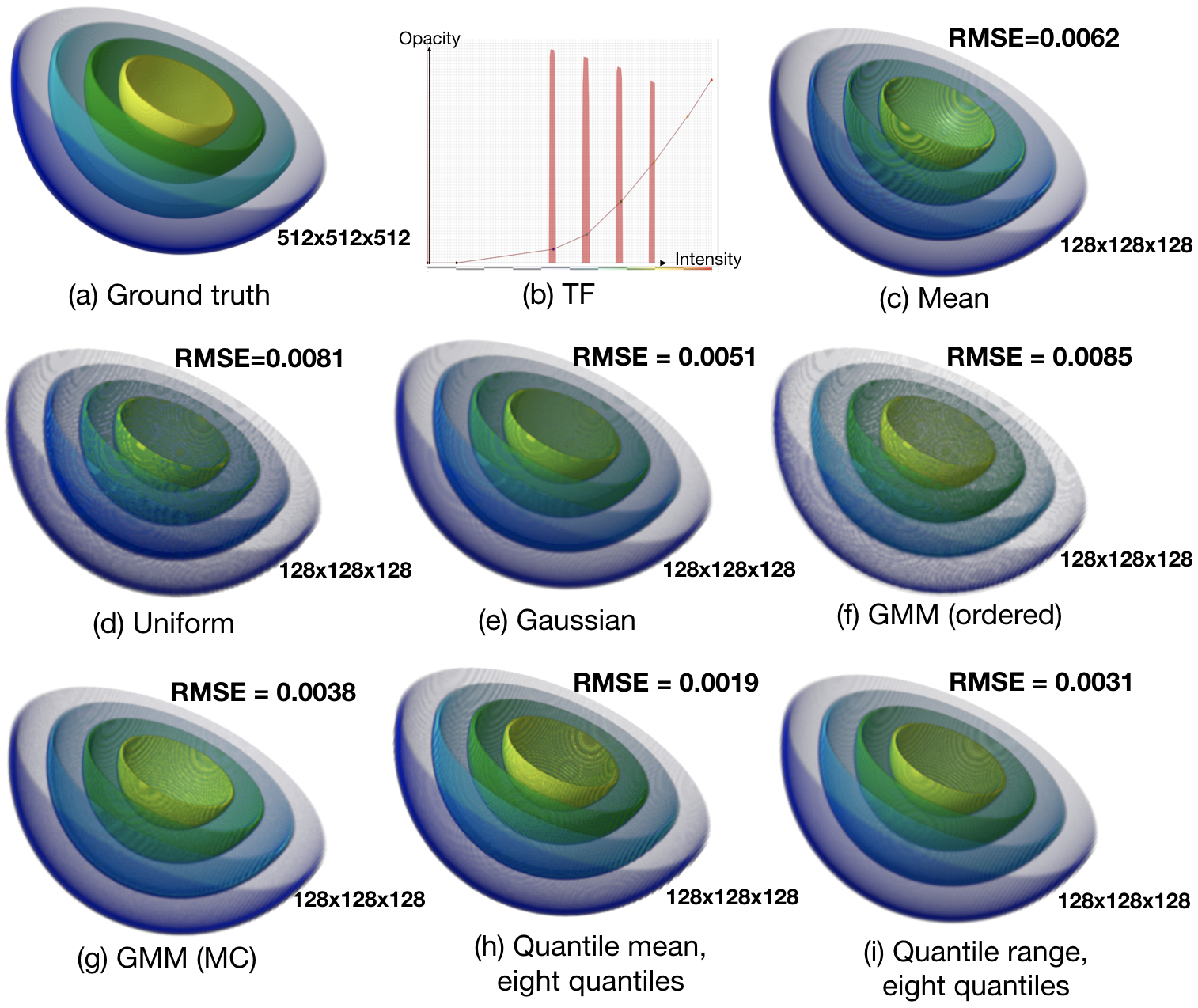}
	\caption{Visualizations of the nested spheres function: All results are rendered with the same TF shown in image (b).} 	\label{fig:nestedSpheres}
\end{figure} 

\paragraph{Downsampled Datasets:} We perform an experiment on a nested spheres dataset similar to the experiment described in Section 4.2 of the previous work on GMM-based DVR~\cite{TA:2012:LLBP}. In \FG{fig:nestedSpheres}, we compare the visualizations for various noise models. For our experiment, we sample the nested spheres function on a high-resolution \ta{$512 \times 512 \times 512$} grid. In the nested spheres function, spheres attain nonzero values on their surface, and the empty spaces between the spheres are assigned zero intensity. We treat the visualization of a high-resolution volume as a reference image (\FG{fig:nestedSpheres}(a)). We set TF (\FG{fig:nestedSpheres}(b)) for the ground truth such that four spheres are assigned distinct colors. The histogram in \FG{fig:nestedSpheres}(b) shows four intensity peaks corresponding to four spheres. The high-resolution nested spheres dataset is partitioned into \ta{$4 \times 4 \times 4$} bricks, and the mean intensity for each brick is stored at a grid vertex of the downsampled version. The mean-field is visualized in \FG{fig:nestedSpheres}(c). \ta{For distribution-based analysis, we store a probability distribution characterizing noise per brick (similar to the hixel idea proposed by Thompson \etal~\cite{Thompson:2013:LDAV}) and apply our DVR framework for visualization of the distribution data. \FG{fig:nestedSpheres}(d-i) visualize the results for various noise models, in which our quantile mean technique exhibits the lowest RMSE. Our closed-form nonparametric models enable us to efficiently generate visualizations without needing to perform MC sampling or screen space integration as in~\cite{TA:2012:LLBP}.} 

\begin{figure} [!ht] 
\centering
\includegraphics[width = 3.3in]{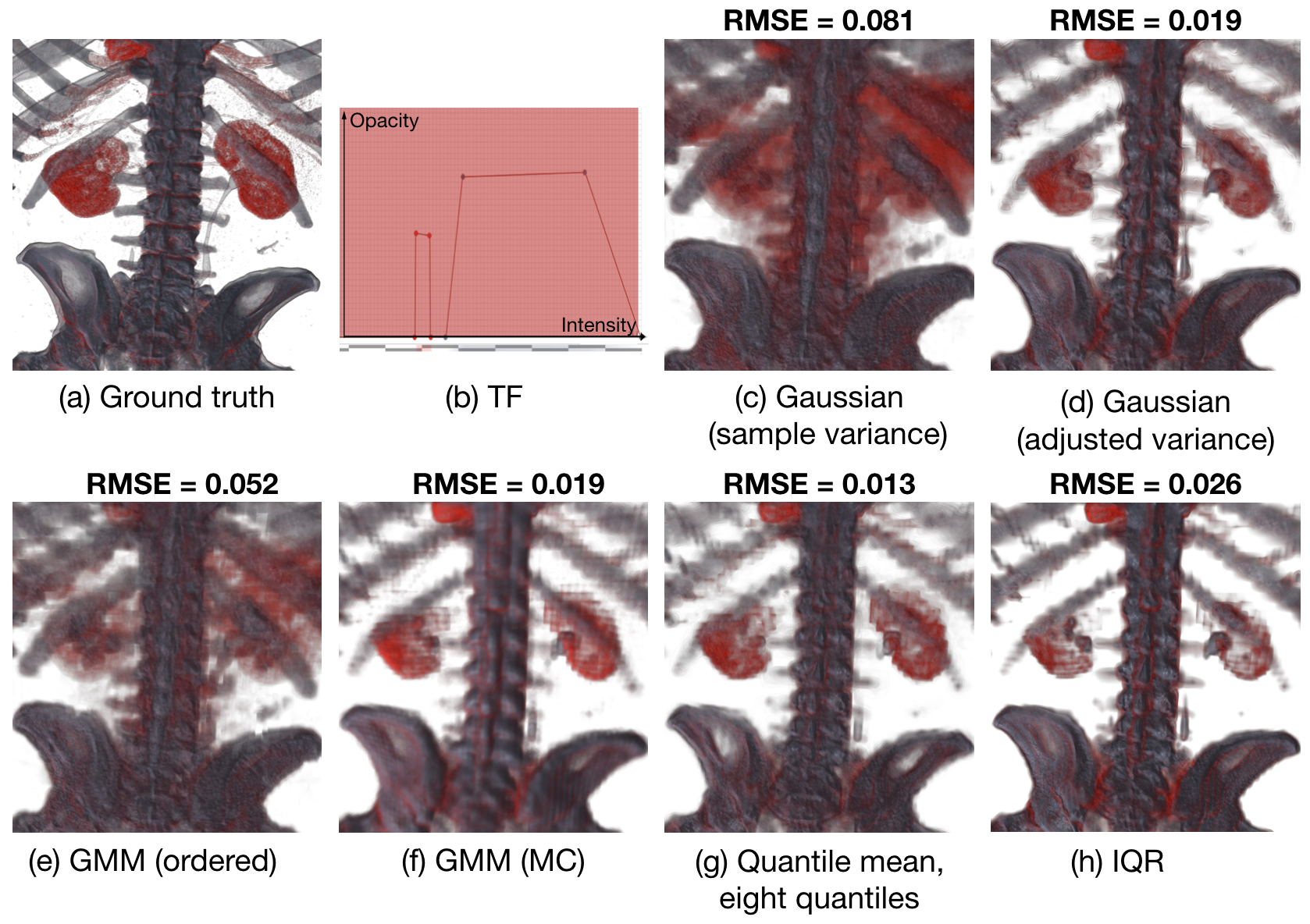}
	\caption{Zoomed-in views for the Osirix OBELIX dataset visualizations in~\FG{fig:teaser}: All results are rendered with the same TF (image (b)). The ground truth in (a) with resolution $512\times512\times1559$ is reduced to $64\times64\times195$, and the reduced data with uncertainty are visualized by applying various noise models in (c)-(h). 
}
\label{fig:obelix}
\end{figure} 

\ta{\FG{fig:teaser}(b) visualizes results similar to \FG{fig:nestedSpheres} on the Osirix OBELIX dataset and classifies bones (gray) and kidneys (red). In \FG{fig:teaser}(b), the mean-field looses a significant amount of information near the kidneys. The visualizations using the uniform and Gaussian noise models result in thicker bone rendering and fail to localize the right-kidney classification. For the zoomed-in visualizations of the Osirix OBELIX dataset in Fig. 9, the RMSE for our quantile mean technique is the smallest among all noise models.}

\paragraph{Parameter Sensitivity:} \tma{We briefly discuss the parameters of statistical models that can potentially influence the quality of visualizations, and hence, RMSE. In the case of our proposed nonparametric models, the reliability of quantile interpolation results is, admittedly, sensitive to the sample size with respect to the number of quantiles (see \FG{fig:quantileInterpolationTechnique} and \FG{fig:tangleQuantileAndSampleNumberEffect}(a)), which in turn depends on the complexity of underlying data distributions. We ensure a sufficient sampling density for the dataset at hand, e.g., $50$ for the tangle dataset and $20$ for the Red Sea eddy simulations, with an empirical approach. Specifically, we keep track of how much visualizations vary with increasing the sample size. We stop when the increase in sample size does not change or affect the quality of visualizations (more details in the supplementary material). We follow the same empirical approach for ensuring sufficient sampling density for all nonparametric statistical visualizations.} 

\tma{In the case of the parametric models, the RMSE is again sensitive to the choice of the parameter values. For example, for the uniform and Gaussian noise models, we estimate the mean and width/variance from the noise samples. Adjusting the Gaussian variance estimated from the noise samples can, however, significantly improve the classification result, as we have demonstrated for the Osirix OBELIX dataset in \FG{fig:obelix}(c) and (d). In the case of our GMM visualizations, we use four Gaussians per mixture since they consume a memory comparable to the quantile interpolation with eight quantiles. Still, the estimation of the parameters, such as the number of Gaussians, regularization value, covariance matrix estimation, and number of MC samples, can be further improved to enhance the quality of the GMM results. Note that the GMMs demand a memory consumption equal to three times the number of Gaussians in a GMM (for storing mean, variance, and weight per Gaussian). In summary, the study regarding the choice of the noise model in the context of DVR and analytical identification of optimal parameter values for data with arbitrary complexity of noise distributions is nontrivial, and we plan to research it in the future.}


\subsection{Application: $\textbf{2}$D TFs and Visualizing Bivariate Data}
We demonstrate our statistical rendering with $2$D TFs using the classic tooth dataset in \FG{fig:tooth}. \FG{fig:tooth}(a) shows the ground truth rendering (top) with a $2$D TF (bottom), where the tooth is classified as four materials: dentine (yellow), tooth-holding material (blue), enamel (red), and root and dentine boundaries (green). \FG{fig:tooth}(b) shows the standard DVR of the bivariate field characterized by mean intensities and the mean gradient magnitudes for data at reduced resolutions. \FG{fig:tooth}(c) shows our statistical rendering using the interpolated distributions of uncertain intensities and gradient magnitudes at the same reduced resolutions as \FG{fig:tooth}(b). When comparing \FG{fig:tooth}(b) and \FG{fig:tooth}(c), our methods show improved recovery of the features in the ground truth, whereas the mean-field visualizations result in poor classifications. 

\begin{figure} [!ht] 
\centering
\includegraphics[width = 2.6in]{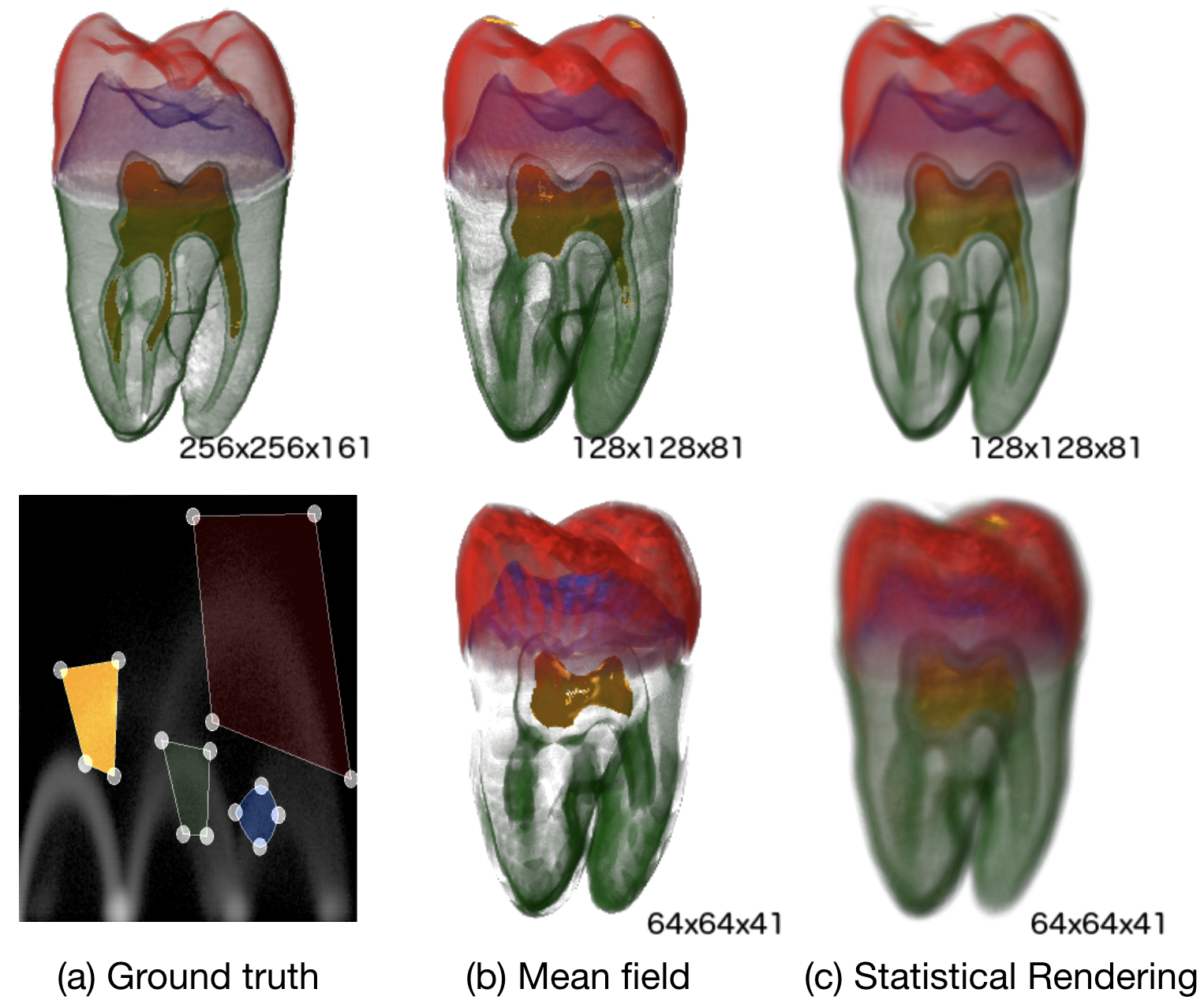}
	\caption{Volume rendering of the tooth dataset with a 2D TF.}
\label{fig:tooth}
\end{figure} 

\begin{figure} [!ht] 
\centering
\includegraphics[width = 3in]{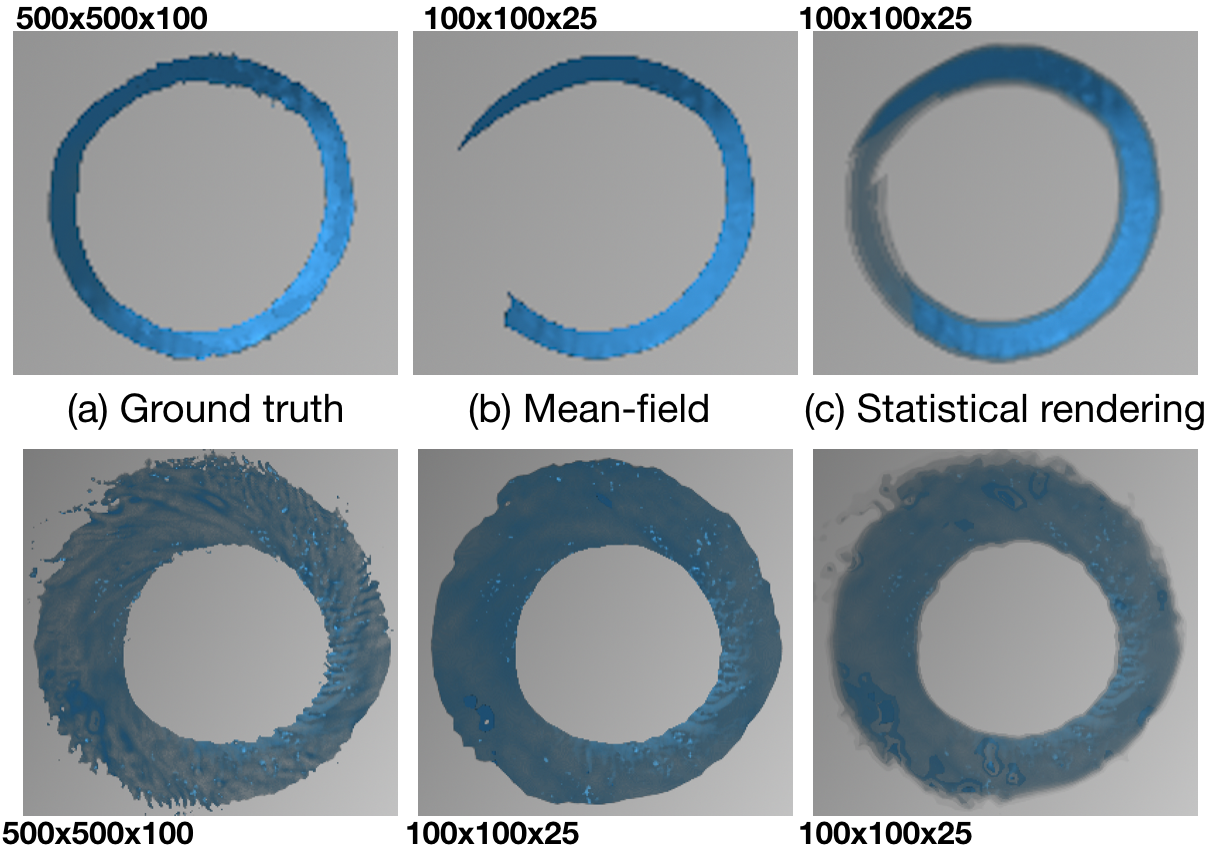}
	\caption{Visualizations of fiber surfaces with $600 <$ Pressure $< 800$ and Water Vapor $=0.011$ in the top row and $1500 <$ Pressure $< 1800$ and Water Vapor $=0.028$ in the bottom row.} 
\label{fig:fiber1}
\end{figure}

Thompson \etal~\cite{Thompson:2013:LDAV} proposed a technique to visualize \textit{fuzzy isosurfaces} of a scalar field in reduced data. The multifield equivalents of isosurfaces are fiber surfaces~\cite{CGF:CGF12636}. In the bivariate case, a fiber surface is a contour defined by a curve that is composed of a number of points (i.e., fibers) in the 2D range space. In the case of reduced bivariate data, our proposed uncertainty-aware 2D TF integration scheme can be applied for the visualization of \textit{fuzzy fiber surfaces}. Specifically, we can model the uncertainties at grid points of the reduced data as the tensor product of two univariate box splines defined by any two uncertain fields. Then, the 2D TF reconstruction scheme introduced in \SC{sec:ProbTF} can be directly applied to the spline-modeled bivariate uncertain field. 

We conducted experiments on a multifield dataset, \textit{Isabel}, a simulation of hurricane Isabel from the West Atlantic region in $2003$. The data dimensions are $500\times500\times100$ with $48$ time steps.~\FG{fig:fiber1}(a) shows the volume rendering of a fiber surface defined in a bivariate field: Pressure and Water Vapor of \textit{Isabel} at time step $18$. The fiber surface is identified by a line in the 2D range space with $600 <$ Pressure $< 800$ and Water Vapor $=0.011$ (the top row of~\FG{fig:fiber1}). When the data are represented as the mean statistics with a resolution of $100\times100\times25$, as shown in ~\FG{fig:fiber1}(b), volume rendering of the reduced fiber surface resulted in a disconnected surface.~\FG{fig:fiber1}(c) shows our statistical rendering of the fiber surface, which preserves the topology of the original surface with "fuzzy" depiction of less certain areas.  
The bottom row in~\FG{fig:fiber1} shows DVR results for another fiber surface similar to the top row results for $1500 <$ Pressure $< 1800$ and Water Vapor $=0.028$.  

 
\section{Conclusion and Future Work}\label{concl}
We expand the spline-based parametric DVR framework~\cite{Entezari2016Statistical} to more flexible nonparametric statistics for visualization of an uncertain scalar field. We leverage the quantile interpolation technique~\cite{TA:1999:Read, TA:2013:HP} for efficient integration of nonparametric PDFs with a DVR framework. \tma{We evaluate our proposed nonparametric statistical models by presenting qualitative and quantitative comparisons with respect to mean-field and parametric statistical models.} We show that the time and memory complexity of our nonparametric DVR framework increases linearly with the number of quantiles used in quantile interpolation. \ta{We demonstrate the application of a previous study~\cite{Entezari2016Statistical} to 2D TFs for improved DVR classification compared to the 2D TF classification with mean statistics.}



\ta{Our approach has a few limitations that we plan to address in subsequent work. In this work, we employ quantile interpolation with even quantile values. We would like to study an efficient use of uneven quantile values for improving DVR classification accuracy, which we briefly discuss in the supplementary material. Next, we would like to investigate methods for efficient implementation of an exponentially complex box-spline framework~\cite{Entezari2016Statistical} (see also the supplementary material) and study their effectiveness in DVR for uncertain data. We plan to generalize our uncertainty-aware 2D TF framework (\SC{sec:gradient}) to TFs with a variable number of dimensions using nonparametric statistics. Finally, we plan to study the nonparametric models for dependent random fields for further improvements in DVR reconstruction accuracy.} 


\acknowledgments{
This work was supported in part by the NSF grant IIS-1617101; the NIH grants P41 GM103545-18 and R24
GM136986; the DOE grant DE-FE0031880; and the Intel
Graphics and Visualization Institutes of XeLLENCE.
}

\bibliographystyle{abbrv-doi}

\bibliography{ms}
\end{document}